\documentclass[prx,twocolumn,floatfix]
{revtex4-2}

\usepackage{graphicx}
\usepackage{amsmath} 
\usepackage{amssymb}
\usepackage{siunitx}

\newcommand{\cut}[1]{}
\newcommand{\be}{\begin{equation}}
\newcommand{\ee}{\end{equation}}
\newcommand{\ba}{\begin{eqnarray}}
\newcommand{\ea}{\end{eqnarray}}
\newcommand{\nn}{\nonumber \\}
\def\ket#1{\left\vert #1 \right\rangle}
\def\bra#1{\left\langle #1 \right\vert}

\begin{document}

\title{Blockchain with proof of quantum work}
\date{\today}

\newcommand{\affildw}{D-Wave Quantum Inc., Burnaby, British Columbia, Canada}

\author{Mohammad H. Amin}
\affiliation{\affildw}
\author{Jack Raymond}
\affiliation{\affildw}
\author{Daniel Kinn}
\affiliation{\affildw}
\author{Gunnar Miller}
\affiliation{\affildw}
\author{Firas Hamze}
\affiliation{\affildw}
\author{Kelsey Hamer}
\affiliation{\affildw}
\author{Joel Pasvolsky}
\affiliation{\affildw}
\author{William Bernoudy}
\affiliation{\affildw}
\author{Andrew D. King}
\affiliation{\affildw}
\author{Samuel Kortas}
\affiliation{\affildw}

\begin{abstract}

We propose a blockchain architecture in which mining requires a quantum computer. The consensus mechanism is based on proof of quantum work, a quantum-enhanced alternative to traditional proof of work that leverages beyond-classical quantum computation to make mining intractable for classical computers. We have refined the blockchain framework to incorporate the probabilistic nature of quantum mechanics, ensuring stability against sampling errors and hardware inaccuracies. To validate our approach, we implemented a prototype blockchain on four D-Wave$^{\rm TM}$ quantum annealing processors geographically distributed within North America, demonstrating stable operation across hundreds of thousands of quantum hashing operations. Our sampling protocol follows the same approach used in the recent demonstration of beyond-classical computation \cite{supremacy}, ensuring that classical computers cannot efficiently perform the same computation task. By replacing classical machines with quantum systems for mining, it is possible to significantly reduce the energy consumption and environmental impact traditionally associated with blockchain mining while providing a quantum-safe layer of security. Beyond serving as a proof of concept for a meaningful application of quantum computing, this work highlights the potential for other near-term quantum computing applications using existing technology.

\end{abstract}

\maketitle

\section{Introduction}

Blockchain technology is widely regarded as one of the most transformative innovations of the 21st century. Its primary purpose is to provide a decentralized and secure method for recording and verifying transactions without the need for a central authority. The concept of blockchain was introduced with the creation of Bitcoin in 2008, as described in the now-famous whitepaper published under the pseudonym Satoshi Nakamoto~\cite{CB1}. Initially developed to enable cryptocurrencies, blockchain technology has since expanded to a diverse range of applications, including decentralized finance (DeFi)~\cite{CB5}, identity verification~\cite{CB4}, supply chain management~\cite{CB2} and healthcare~\cite{CB3}. Key characteristics of blockchains include transparency (all transactions are visible to participants in the network) and resistance to tampering (it is extremely challenging for malicious actors to modify data on the blockchain, so transactions are practically immutable)~\cite{CB6,CB7,CB8}. These features have made blockchain technology a cornerstone of the shift toward Web 3.0, the next generation of the internet \cite{CB9,Web3}.

At its core, a blockchain is a distributed ledger that operates across a network of computers, where data is stored in ``blocks" that are cryptographically linked to form a sequential chain. This architecture eliminates the need for central intermediaries, enabling trustless interactions between parties. The integrity of the data is secured through consensus mechanisms, the process by which blockchain networks agree on the validity of transactions and ensure the synchronization of their distributed ledgers. In the absence of a central authority, consensus mechanisms enable trust and cooperation among participants in the network. The choice of consensus mechanism impacts a blockchain's security, scalability, energy efficiency, and decentralization, making it a key component of the network's design. These mechanisms ensure that no single entity can control the ledger, thus preserving its decentralized nature. However, they also present challenges.

Traditional blockchains use consensus algorithms like proof of work (PoW) and proof of stake (PoS) to validate transactions. PoW relies upon the significant computational effort to generate valid blocks. Distributed computational power ensures that no user can control the blockchain evolution. Introducing invalid transactions, such as double-spending, requires of the order of 50\% of the community compute power, which is financially prohibitive due to the substantial equipment and electricity costs. While PoW effectively enhances network security, it also leads to significant energy consumption. Bitcoin mining alone consumed an estimated 175.87 terawatt-hours (TWh) in 2024, surpassing the yearly electricity usage of entire countries, including Sweden \cite{BitCoinCons, FinTimes}. This high energy consumption raises environmental concerns.

PoS addresses this issue by selecting validators based on the amount of cryptocurrency each holds and is willing to stake as collateral, incentivizing honest behavior by exposing their holdings to potential loss in case of misconduct. However, a major drawback of PoS is the potential for wealth concentration, as those with larger holdings have a greater chance of being selected as validators, further increasing their wealth over time. This can reduce the network's decentralization, a core principle of blockchain technology. Other mechanisms, such as delegated proof of stake (DPoS) \cite{CB11} and practical Byzantine fault tolerance (PBFT) \cite{CB12}, offer alternative approaches tailored to specific use cases.

As blockchain technology advances, researchers and technologists are investigating innovative approaches to overcome its limitations and ensure its long-term viability. One exciting avenue is the integration of quantum computing with blockchain systems, a combination that could transform the field by offering unmatched security, scalability, and computational power. This article introduces a blockchain architecture that incorporates a proof of quantum work (PoQ) consensus mechanism, where quantum computers are required to validate transactions. The central concept is to harness beyond-classical quantum computation to mine problems intractable to classical computers. Quantum supremacy is a term introduced by John Preskill \cite{Preskill2012} that refers to a scenario in which a quantum computer performs a task that classical computers cannot accomplish with reasonable resources. While there could be various ways to leverage this quantum advantage within a blockchain, our proposal builds on Bitcoin's architecture with minimal modifications, taking advantage of its proven reliability and robustness. 

Unlike some previous quantum blockchain proposals \cite{QB1,QB2,QB3,QB4,QB5,QB6,QB7,QB8,QB9,QB10}, which often assume access to fault-tolerant quantum computing or quantum teleportation, our approach integrates both classical and quantum computation to leverage quantum advantage with existing and near-term QPUs. Our proposal is also different from blockchains and hashes that have been proposed to exploit the potential of near-term boson-sampling \cite{nikolopoulos_cryptographic_2019, shi_quantum_2022, singh_proof--work_2025}. We introduce a hierarchy of practical witness methods, new consensus mechanisms robust to noise, and implement a proof-of-concept demonstration using a distributed network of quantum computers.

Although claims of beyond-classical computation have been made in Boson sampling, these have been challenged~\cite{madsen_quantum_2022, oh_classical_2024}. Currently, Ref.~\cite{supremacy} is the only demonstration of beyond-classical computation not based on random-circuit sampling~\cite{Google2024,Chinese2025}, and thus the only demonstration of quantum advantage relevant to this work (see Appendix \ref{RCS}). The unitary dynamics used for hashing remains intractable despite recent developments in algorithmic methods~\cite{tindall_dynamics_2025, mauron_challenging_2025, king_comment_2025}.

Using a small-scale prototype, we demonstrate this blockchain's functionality on four generally-available D-Wave Advantage$^{\rm TM}$ and Advantage2$^{\rm TM}$ quantum processing units (QPUs) at different geographic locations across North America. By carefully accounting for the probabilistic nature of quantum computation and potential errors within Bitcoin's architecture, we achieve stable operation across thousands of mining processes distributed among hundreds of stakeholders. This marks the first successful demonstration of an important real-world application that directly leverages beyond-classical computation, operating seamlessly across an internationally distributed network of quantum computers.

\section{Classical Blockchain Architecture}

Before introducing any quantum concepts, we first describe how standard blockchain technology works, using Bitcoin's architecture as a prime example. At the heart of Bitcoin is its distributed ledger, a decentralized database that records all transactions. This ledger is maintained across a network of computers, ensuring both transparency and security. Each node on the network stores a complete copy of the blockchain, and updates are synchronized through consensus mechanisms. These mechanisms ensure that only valid transactions are added to the blockchain, making it resistant to tampering or fraud.

\begin{figure} 
\includegraphics[width=\linewidth]{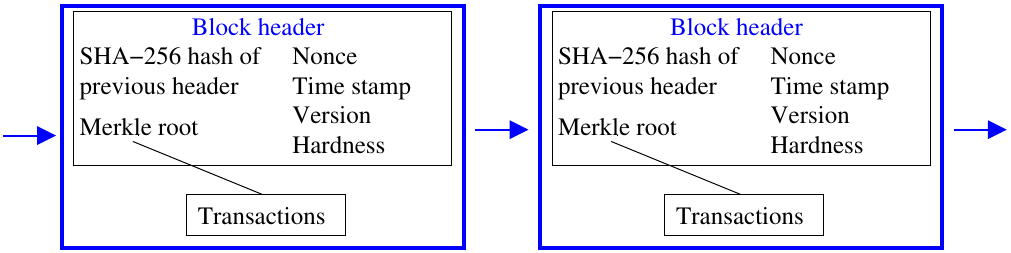}
\caption{\label{block-structure} Illustration of Bitcoin blocks. The Bitcoin block structure is sufficient for our proof-of-concept implementation; modifications may be useful for mitigation of weaknesses as discussed in Appendix \ref{attacks}.}
\end{figure}

Transactions are recorded in blocks linked sequentially in a chain, forming a continuous and immutable ledger. A Bitcoin block consists of a block header and a list of transactions (see Figure~\ref{block-structure} for illustration). The block header acts as a metadata section and contains essential elements such as a timestamp, a Merkle root (a hierarchical summary of all transactions in the block), the nonce, and the hash of the previous block. A hash is a unique, fixed-length binary string generated by processing data through a cryptographic algorithm. It acts as a digital fingerprint, ensuring data integrity by making even the slightest change to the input data result in a completely different hash. The nonce is an arbitrary 32-bit number adjusted by the miners to solve the cryptographic puzzle required for block validation. Miners achieve this by repeatedly combining the transaction data with the nonce and passing it through a classical hashing algorithm, which generates an ${\cal N_H}$-bit hash ${\cal H}$ (${\cal N_H} = 256$ for Bitcoin). Adding a block to the blockchain begins with miners competing to find a nonce that produces a hash value below a specific target ${\cal H} < {\cal T}_n$, with $n$ indexing the block. The inequality requires that the first ${\cal N}_{\rm zeros}$ bits of the hash be zero, where
\be
{\cal N}_{\rm zeros} = \log_2 \Bigg( {2^{\cal N_H} \over {\cal T}_n + 1} \Bigg).
\ee
Due to the irreversibility of the hash function, finding the correct hash can only be achieved through brute-force search, forcing miners to test countless nonce values. This process, known as ``mining," relies on a trial-and-error approach, requiring intensive computational effort to solve the cryptographic puzzles that underpin PoW. In Bitcoin, the network dynamically adjusts the difficulty every 2016 blocks (approximately every two weeks) by redefining the target ${\cal T}_n$, to ensure that the average time to mine a block remains close to 10 minutes.  Once a miner successfully solves the puzzle, they broadcast the solution to the network. Other nodes then validate the solution by verifying the hash and ensuring the block's transactions are legitimate. Upon successful validation, the block is added to the chain, and the miner is rewarded.

The reward for mining a block consists of two components: the block reward and transaction fees. The block reward is a fixed amount of Bitcoin awarded to the miner. This reward halves approximately every four years in an event called the ``halving." In addition to the block reward, miners collect fees from transactions included in the block, incentivizing them to prioritize transactions with higher fees. This dual reward structure ensures the security and continuity of the Bitcoin network while gradually reducing the issuance of new Bitcoin, contributing to its scarcity.

A situation known as a ``fork" occurs when two miners solve a block at nearly the same time, resulting in the blockchain temporarily splitting into two competing branches. Both branches propagate through the network, and nodes may accept one branch over the other depending on which block they receive first. The fork is resolved when miners adding new blocks to one of the branches make it longer than the other. Bitcoin follows the ``strongest chain rule", meaning the branch with the highest total computational work (often called Chainwork) is recognized as the valid chain, while the shorter branch is discarded. This process ensures network consensus, but can result in transaction delays and unrewarded work. It is therefore essential to understand how the strongest chain is determined.

In Bitcoin, a set of valid proposed blocks forms a directed tree, rooted in a special genesis block, the first block introduced by Satoshi Nakamoto in 2009. Any path including an invalid block is to be ignored by stakeholders. Each node on the tree has an associated work, which is a sum of all work completed from the genesis block to the node (along a unique directed path). For the $n$-th block in the chain, the work is defined as 
\be
{\cal W}_n = 2^{{\cal N}_{\rm zeros}}.
\ee
Note, the $n$-dependence of ${\cal N}_{\rm zeros}$ is omitted as an abbreviation.
It represents the number of trials and errors, required to find an acceptable hash, ${\cal H} < {\cal T}_n$. Therefore, ${\cal W}_n$ quantifies the mining work, which goes up as the hash-hardness grows. The Chainwork is for practical purposes monotonic in the length of the chain. The node with the most Chainwork defines a unique path from the genesis node called the strongest chain. The total Chainwork is therefore defined as
\be
{\rm Chainwork} = \sum_{n \, \in \, {\rm chain}} {\cal W}_n
\ee
Chainwork is calculated by the stakeholders for all existing chains and the chain with largest Chainwork is selected for adding the next block and the rest are discarded. The miners responsible for the discarded blocks do not receive rewards, while the miner who successfully extends the longest chain is rewarded with both block rewards and transaction fees from the new block. Transactions from the discarded branch are returned to the memory pool (mempool), where they await inclusion in a future block. This mechanism ensures that consensus is maintained across the network, although it may introduce temporary delays in transaction confirmation. 
On a strongest chain, blocks closer to the genesis block are increasingly likely to remain unchanged as broadcasts continue. If we define a small cut-off in probability, we might say blocks become (for practical purposes) immutable, and define the transaction delay accordingly.
In the case of Bitcoin it is only a few blocks, assuming  good network synchronization. Miners are incentivized to maintain synchronization, to minimize the risk of pursuing wasteful computations, so that forks are merely a rare inconvenience and a delay of no more than 2 blocks is typical. Forks may also happen due to software updates and change of rules, which may lead to temporary (soft-fork) or permanent (hard-fork) splitting of the blockchain.

We highlight three features required for viability of a blockchain:  1. The number of blocks in the strongest chain relative to the number of honestly-mined blocks broadcast should be a large finite fraction (the efficiency). 2. A stakeholder should have confidence that that blocks buried to some small finite depth on the strongest chain are immutable with high probability. The ratio depth/efficiency dictates the expected number of honestly-mined blocks that have to be broadcast before the most recently broadcast transactions can be considered reliably processed. 3. Consensus mechanisms should protect against disruptions, and prevent rewards from being obtained (on average) by incomplete or fraudulent work. These are necessary requirements for guaranteeing confidence, and maintaining an incentive to mine according to protocol rules. 

\section{Classical Hashing}

Hash algorithms are widely used for various purposes, including verifying data integrity, securely storing passwords (by hashing them instead of storing them in plain text), creating hash tables in data structures, and enabling cryptocurrencies. These algorithms are a fundamental tool in secure communications. It should be clear by now that cryptographic hashing plays a crucial role in blockchain technology. In essence, hashing is the deterministic process of transforming a message or data of any length into a unique fixed-size bit string, known as the hash. The primary distinction between cryptographic hashing and encryption is that hashing is a one-way transformation designed to generate a unique, irreversible output, whereas encryption is a two-way process intended to securely encode data while allowing the original message to be recovered with a decryption key.

Cryptographic hashing has several important properties that make it suitable for various applications, including the following:

\begin{enumerate}

\item {\bf Fixed output size:} The hash function should generate a fixed-size output, regardless of the size of the input.

\item {\bf Collision resistance:} It should be very unlikely that two different inputs produce the same hash value. 

\item {\bf Avalanche effect:} A small change in the input data should result in a significantly different hash value (similar inputs should not produce similar hash values).

\item {\bf Uniform distribution:} The hash values should be uniformly distributed across the output space to avoid clustering, which can lead to collisions.

\item {\bf Pre-image resistance:} It should be computationally infeasible to infer anything about the original input data from the hash. Given an input, it should be infeasible to find a second input yielding the same hash. 

\item {\bf Deterministic:} For the same input, a hash function must always produce the same output.

\end{enumerate}

Commonly used hash algorithms include MD5, SHA-1, SHA-2, and the SHA-3 family, all introduced by the National Institute of Standards and Technology (NIST) \cite{NISTSHS}. Among these, SHA-256 (Secure Hash Algorithm 256-bit) is the standard hash function used in Bitcoin and many other blockchain applications. It works by processing input data through a series of mathematical operations, including bitwise logical operations, modular arithmetic, and compression functions. The input message is first padded to a specific length and divided into 512-bit blocks. Each block is then processed through 64 rounds of mathematical transformations, which involve constants and logical functions, resulting in a fixed 256-bit hash output. This hash is unique to the input data and ensures that even a tiny change to the input produces a drastically different hash, making it highly secure and resistant to collision attacks. Beyond SHA-256, the SHA-2 family includes variants like SHA-384 and SHA-512, which offer larger output sizes for even greater security, and the newer SHA-3 family provides enhanced resistance to potential cryptographic attacks.

\section{Hybrid Classical-Quantum Hashing}

\begin{figure*}
\begin{center}
\includegraphics[width=0.8\linewidth]{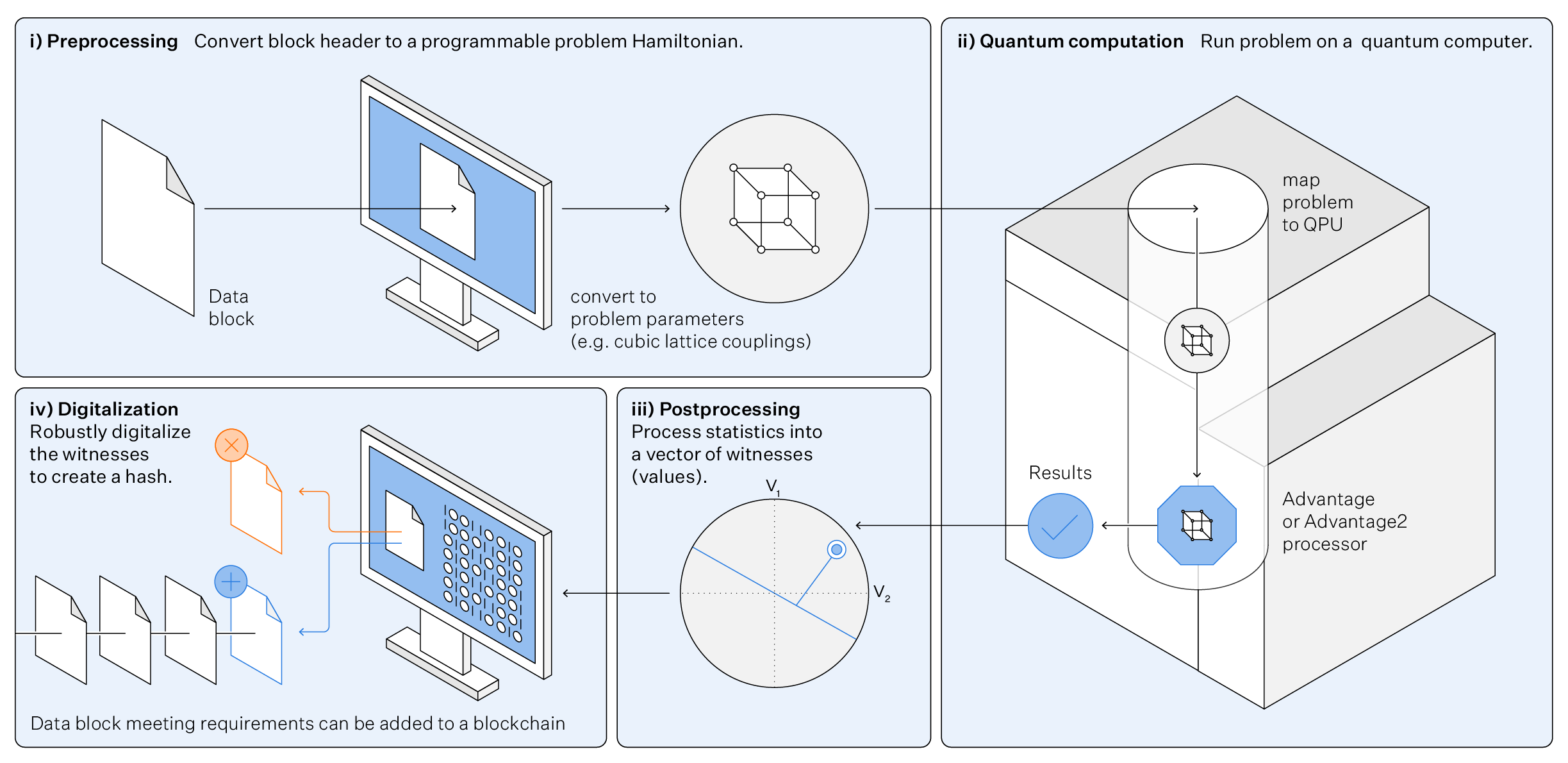}
\caption{\label{fig:cycle} Illustration of the hybrid classical-quantum hash generation and its use as proof of work for block security.}
\end{center}
\end{figure*}

Various quantum hashing algorithms have been proposed, which can generally be categorized into two types. The first type produces a quantum state as the output, requiring communication and authentication to be quantum mechanical \cite{QQH1,QQH2,QQH3,QQH4,QQH5,QQH6,QQH7,QQH8}. The second type leverages quantum mechanics to generate a classical hash value, which can then be communicated and processed using classical means \cite{QCH1,QCH2,QCH3}. In this paper, we focus on the latter.

The main challenge in quantum hashing is that quantum mechanics is inherently probabilistic, while standard hash functions are deterministic. However, although the outcome of any single quantum measurement is probabilistic, the probability distribution is a deterministic function of the system parameters. Similarly, all expectation values derived from the distribution are deterministic functions of these parameters. This allows a quantum hash function to be uniquely defined based on a quantum probability distribution and its corresponding expectation values. In practice, however, accessing this probability distribution is possible only through sampling, so that accuracy is limited. This process is inevitably subject to inaccuracies due to sampling and hardware-induced errors. These factors can reduce the precision and reliability of the quantum hash function.  

A quantum hash function must satisfy the first five classical requirements of hash functions, outlined in the previous section. The final requirement, being deterministic, is not possible and must be replaced with two new quantum-specific requirements:

\begin{enumerate}
\item[6.] {\bf Probabilistic:} For the same input, a quantum hash function has a nonzero likelihood of producing different outputs.

\item[7.] {\bf Spoof resistance:} The hash must be infeasible to replicate using classical computers with realistic resources.
\end{enumerate}

\noindent The last requirement is essential when the hash is used in a blockchain with a PoQ-based consensus mechanism. It ensures that only quantum computers can perform mining, thereby preventing energy-intensive classical mining. The probabilistic nature of quantum hash values must be carefully considered in the design and application of hash functions, particularly in systems like blockchain, where reliable behavior is essential. This issue, along with potential solutions, will be discussed in greater detail in subsequent sections.
Rather than requiring our quantum evolution to provably guarantee cryptographic properties, it is convenient to exploit a hybrid approach with one classical cryptographic hash used to define the quantum-computation parameters.
Since existing classical cryptographic hashes already fulfill the first five classical requirements the focus can shift to achieving the last requirement.

We start by introducing the general concept without delving into specific implementation details. Let $\mathcal M$ represent a message of arbitrary length. (For clarity and consistency, we use calligraphic symbols for blockchain-related quantities, while standard symbols are reserved for all other variables, including those characterizing the quantum evolution.)  A hash function is defined as a unique transformation of $\mathcal M$ into a hash value ${\cal H}(\mathcal M)$, with a fixed length of $\mathcal  N_\mathcal{H}$ bits. This transformation is performed in four steps (see Fig.~\ref{fig:cycle} for visualization):

\begin{enumerate}

\item[i.]  {\bf Preprocessing:} Transform message $\mathcal M$ into a set of random parameters $\Theta$
\be
\Theta = {\rm Random}(\mathcal M).
\ee

\item[ii.] {\bf Unitary evolution:} Use $\Theta$ to define a unitary evolution $U_\theta$ for an $N_Q$-qubit quantum system with initial state $|\psi_0 \rangle$ such that:
\be
|\psi_\mathcal{M}\rangle = U_\Theta |\psi_0 \rangle,
\ee
where $|\psi_\mathcal{M} \rangle$ is the final state with its corresponding pure state density matrix $\rho_\mathcal{M} = |\psi_\mathcal{M} \rangle \langle \psi_\mathcal{M} |$. The final state may be described in a compressed form by a set of expected (measured) observables.

\item[iii.] {\bf Postprocessing:} Using samples obtained from multiple measurements, compute a vector $W(\rho_\mathcal{M})$ consisting of observables (witnesses) $W_\alpha$.

\item[iv.] {\bf Digitalization:} Discretize witnesses into binary numbers, e.g.,
\be \label{Halpha}
{\cal H}_\alpha = \theta(W_\alpha {-} W_{0\alpha}) = \left\{ \begin{array}{cc}
    0, & \quad W_\alpha < W_{0\alpha} \\
    1, & \quad W_\alpha \geq W_{0\alpha}
\end{array}
\right. ,
\ee
where $\theta(x)$ is the Heaviside step-function and $W_{0\alpha}$ is a threshold selected depending on the definition of $W_{\alpha}$. 
The hash is a sequence of the binaries:
\be
{\cal H}(\mathcal M) = {\cal H}_1 {\cal H}_2 \dots {\cal H_{N_H}}.
\ee
\end{enumerate}
 
In the preprocessing step, the message $\mathcal{M}$ is transformed into a set of parameters $\Theta$ that govern the quantum operation. These parameters may include Hamiltonian variables, evolution times, a circuit, or other relevant quantities. If the size of $\Theta$ remains fixed regardless of the length of $\mathcal M$, the preprocessing step effectively acts as a classical hash. Steps (iii-iv) can also be parameterized as a cryptographic function of $\Theta$ and the message. Standard cryptographic hashing algorithms can be employed at this stage to ensure, with care, that the first five criteria outlined in the previous section are met. This can be achieved by first hashing the message using a classical algorithm, such as SHA-256, and then using the resulting hash as a seed for a pseudo-random number generator to generate 
$\Theta$ and other hashing parameters preventing reversibility. This can preserve the essential properties of the classical hash in $\Theta$, which in turn defines the quantum evolution in the next step.

The quantum operation must be sufficiently complex to satisfy the seventh requirement, ensuring that the hash cannot be spoofed by classical computers with feasible resources. Such complexity can be achieved through a continuous-time evolution, sequence of quantum gate operations, or combination of digital and analog operations. Information from the resulting density matrix is extracted through measurements and summarized in a vector of witnesses $W(\rho_\mathcal{M})$. It should be noted that some classical properties, such as collision resistance and the avalanche effect, do not necessarily carry over automatically. For example, if quantum evolution maps all $\Theta$ to the same hash value, these properties would be lost. One should therefore be careful in the design of the quantum evolution and the witness vector.
The definition of witnesses and the quality of the resulting hash ultimately depend on the quantum hardware's capability to execute complex quantum evolutions and its measurement capabilities, as we discuss next. 

The 4-stage process outlined includes both classical and quantum computation. Although the classical part likely includes non-trivial operations such as cryptographic hashing, it is straightforward to demonstrate the quantum computation costs (time, energy or dollars) are significantly larger than the combination of classical operations, as we later quantify in context of our demonstration.

If hardware limitations constrain qubit measurement to a single basis, which we herein refer to as the $z$-basis (associated with the $\sigma^z_i$ Pauli matrices), the only accessible information is the diagonal elements of the density matrix in this basis. This corresponds to the probability distribution:
\be
P_\mathcal{M} = {\rm diagonal}(\rho_\mathcal{M}).
\ee
The postprocessing step then aims to convert $P_{\mathcal M}$ into a unique bit string of predetermined length $\mathcal{N}_\mathcal{H}$, ensuring robustness to uncertainty in the experiment whilst capturing non-trivial statistical properties. In other words, the resulting quantum hash is a fixed-length bit string that uniquely identifies the probability distribution $P_\mathcal{M}$. A complete characterization of $P_\mathcal{M}$ for a system with $N_Q$ qubits requires $2^{N_Q} {-} 1$ real parameters. This characterization can be achieved by determining the probability of each state in the computational basis or by calculating all expectation values of single- and multi-qubit operators up to order $N_Q$, i.e., $\langle \sigma^z_i \rangle$, $\langle \sigma^z_i \sigma^z_j \rangle$, $\langle \sigma^z_i \sigma^z_j \sigma^z_k \rangle$, etc. In practice, for quantum systems with local interactions, it is often sufficient to consider a limited number of dominant k-local expectations within the correlation length to characterize the distribution with acceptable accuracy. This simplification significantly reduces the complexity of the problem, particularly when the primary goal is to ensure collision resistance in the resulting quantum hash.

Let $V$ represent a vector of length $N_V$ of the dominant expectation values
\be
V = [...,\langle \sigma^z_i \rangle,..., \langle \sigma^z_i \sigma^z_j \rangle,... \langle \sigma^z_i \sigma^z_j \sigma^z_k \rangle, ...]^T
\ee
Each element of the feature vector $V$ is a real number $-1 {\leq} V_\alpha {\leq} 1$. 
To generate an ${\cal N_H}$-bit hash we create ${\cal N_H}$ witnesses as linear combinations of $V_\beta$
\be
W_{\alpha} = {1\over N_V} \sum_{\beta=1}^{N_V} G_{\alpha\beta} \, V_{\beta}
\ee
where $G$ is an ${\cal N_H}{\times} N_V$ matrix with $O(1)$ elements such that $-1 {\leq} W_\alpha {\leq} 1$. We turn each element of the witness vector $W$ into a binary number using \eqref{Halpha}. Each equation $W_\alpha = W_{0\alpha}$ defines a hyperplane in the vector space made of $V$'s and each hash bit $\mathcal{H}_\alpha$ indicates on which side of the hyperplane the observed $V$ lies. 
The threshold value $W_{0\alpha}$ can be taken to be zero or as some simple function of the experimental parameters.
Robustness of the hash bits towards errors and spoofing depends on the choice of the hyperplanes (see Appendix \ref{HyperplaneDistributionMitigationsSection}). The signs on hyperplane projections can be chosen uniformly at random as a cryptographic function of the message (as in our proof of concept) to guarantee a uniform distribution of bit strings that is not a reversible function of other experimental parameters, providing pre-image protection.

When qubit readout is performed in a single basis, it is, in principle, possible to identify a classical system that can replicate the same distribution without entanglement. While finding such a classical system may be practically infeasible due to beyond-classical computation, its theoretical existence cannot be ruled out. Indeed, any proof of entanglement, whether through an entanglement measure or a witness, requires access to information beyond single-basis measurements of the quantum states. This additional information could be obtained by performing measurements in multiple bases or by examining the system's response to perturbations (e.g., susceptibility measurements). In the absence of more complex measurement, one may resort to susceptibility measurements to protect further against classical spoofing.

Multi-basis measurements and shadow tomography allow additional near-term flexibility in the hash definition. In particular we note that entanglement witnesses might be used, with thresholds in/at the entangled regime. Since classical probabilistic models do not allow for beyond-threshold statistics, this makes the experimental outcome difficult to simulate by any classical models. These methods are described in detail in Appendix~\ref{WitnessConstructionBeyondTheComputationalBasis}.

\section{Quantum Blockchain with Probabilistic Hashing}
\label{QuantumBlockchainWithProbabilisticHashSection}
As previously emphasized, an inherent feature of any quantum mechanical hash generation is its probabilistic nature. Therefore, it is crucial to ensure that the blockchain remains reliable despite this uncertainty. Two primary sources of error can affect the estimation of observables used for hash generation: sampling error and hardware inaccuracies. Sampling error can be reduced by increasing the number of samples, while systematic hardware inaccuracies pose less of a problem if they remain consistent across both the miner and validator quantum devices. However, random errors and variations between different QPUs can introduce discrepancies, potentially leading to disagreements between miners and validators. To address this issue, we introduce two key modifications to the standard blockchain algorithm: probabilistic validation and a revised definition of the strongest chain.

\subsection{Probabilistic Validation}

We now develop a model of probabilistic validation assuming independent and Gaussian distributed witness statistics, which is consistent with witnesses being determined by a sum of many weakly correlated statistics, and proves to be a good approximation in demonstrations as shown later. We introduce a confidence measure that quantifies the accuracy of hash generation and validation relative to an ideal, error-free scenario. The proposed confidence measure can serve as a principled heuristic for practical applications, independent of the underlying assumptions.

Let us assume that the experimentally extracted $W_\alpha$ in Eq.~\eqref{Halpha} follows a Gaussian distribution with a mean of $\bar W_\alpha$ and standard deviation $\delta W_\alpha$ as
\be
P(W_\alpha) = {1 \over \sqrt{2 \pi} \delta W_\alpha} \exp\left({-\frac{( W_\alpha - \bar W_\alpha)^2}{2\delta W_\alpha^2}}\right).
\ee
In the absence of any systematic error, $\bar W_\alpha$ should be the same as the theoretical value. While the exact theoretical values are inaccessible due to the computational infeasibility of solving the Schr\"odinger equation at large scales, $\bar W_\alpha$ and $\delta W_\alpha$ can be estimated through statistical averaging. For a single run of the experiment, however, one only has access to the extracted $W_\alpha$. Assuming that $\bar W_\alpha$ follows the same Gaussian distribution centered around $W_\alpha$, one can estimate the probability of error in the generated hash bit $\mathcal H_\alpha$:
\ba
\varepsilon_\alpha &=& P \big[(W_\alpha {-} W_{0\alpha})(\bar W_\alpha {-} W_{0\alpha}) < 0 \big] \nn
&=& {1 \over \sqrt{\pi}}  \int_{d_\alpha}^\infty e^{-x^2} dx = {1\over 2} [1- {\rm erf}(d_\alpha)]
\ea
where ${\rm erf}(x)$ is the error function and 
\be
d_\alpha = {|W_\alpha - W_{0\alpha}| \over \sqrt{2} \, \delta W_\alpha} 
\ee
represents a dimensionless distance from the corresponding hyperplane. As expected, $\varepsilon_\alpha \to 0$ for $d_\alpha \gg 1$ and $\varepsilon_\alpha \to 0.5$ for $d_\alpha \to 0$. It is more convenient to work with confidence values defined as ${\cal P}_\alpha \equiv 1- \varepsilon_\alpha$, instead of error rates. Assuming that errors in witnesses are uncorrelated, the miner's confidence in the generated ${\cal N_H}$-bit hash value is given by
\be
{\cal P}_{\rm miner}  = \prod_{\alpha = 1}^{N_H} {\cal P}_\alpha.
\ee
If for all hash bits the distances from the hyperplanes are large, the confidence value will be close to one. In contrast, if any distance $d_\alpha$ is sufficiently small, confidence value for the corresponding bit approaches 1/2, reducing the overall confidence value. In that case, the generated hash may be rejected by the validators. The number of such uncertain bits is given by
\be
{\cal N}_{\rm miner} = - \log_2 {\cal P}_{\rm miner}.
\ee

The validator performs the same calculations to determine 
$W_\alpha$ and $d_\alpha$ and the hash bit $\mathcal H_\alpha$, which may or may not match the proposed hash bit $\mathcal H'_\alpha$. The validator can calculate their confidence value on the received hash as:
\be
{\cal P}_{\rm validator} = \prod_{\alpha = 1}^{N_H} \big[ {\cal P}_\alpha \delta_{\mathcal H_\alpha,\mathcal H'_\alpha} + (1 -{\cal P}_\alpha)(1 {-} \delta_{\mathcal H_\alpha, \mathcal H'_\alpha}) \big],
\ee
where $\delta_{a,b}$ is the Kronecker delta-function. If all hash bits match, the validator's confidence in the received hash bit is equal to their confidence in their own bit. However, if a single bit is incorrect, the confidence value decreases by a factor of $(1{-}{\cal P}_\alpha)/{\cal P}_\alpha$. This would significantly lower ${\cal P}_{\rm validator}$ if the incorrect bit is one with high confidence value, where ${\cal P}_\alpha \approx 1$. Conversely, if the incorrect hash bit has ${\cal P}_\alpha \approx 1/2$, the reduction in the overall confidence value is negligible. We can turn  ${\cal P}_{\rm validator}$ into a number of bits in a similar way as for the miner:
\be
{\cal N}_{\rm validator} = - \log_2 {\cal P}_{\rm validator}.
\ee
The interpretation of ${\cal N}_{\rm validator}$ is slightly different. If discrepancies between the hashes occur only in bits with a small distance $d_\alpha$, then 
${\cal N}_{\rm validator}$ represents the count of those uncertain bits. However, even a single bit discrepancy outside this set can cause ${\cal P}_{\rm validator}$ to vanish, leading 
${\cal N}_{\rm validator}$ to diverge, effectively rendering all hash bits uncertain. 

The simplest way of implementing probabilistic validation is to set a threshold ${\cal N}_{\max}$ for the allowable discrepancy between the received hash and the one generated by the validator. The validator would then accept the block if ${\cal N}_{\rm validator} < {\cal N}_{\max}$ and reject it otherwise. The stability of the chain can be further enhanced if miners and validators agree on the same ${\cal N}_{\max}$. Provided broadcasting has reasonable cost, the miner will refrain from proposing a block if ${\cal N}_{\rm miner} > {\cal N}_{\max}$ because it is very likely to be rejected by the validators. The advantage of using ${\cal N}_{\rm miner}$ and ${\cal N}_{\rm validator}$ instead of the Hamming distance between the hashes is that it prevents malicious actors from exploiting the relaxed requirements to generate invalid hashes. This is because only hash bits with a small distance $d_\alpha$ are permitted to differ, and calculating $d_\alpha$ can only be achieved accurately using a quantum computer, making it challenging for attackers to identify uncertain bits and manipulate specific hash bits confidently. It also deters a miner from generating the hash with less computational effort, such as by producing a hash with fewer leading zeros and randomly flipping the remaining bits. In practice one needs to fine-tune ${\cal N}_{\max}$ to achieve maximum stability of the blockchain.

Note that $\delta W_\alpha$ and $\mathcal{N}_{\max}(\mathcal{N}_{\mathcal H})$ (a monotonically increasing function of the block length), should be determined one-time as a public rule of the protocol. Targeting some operational delay, efficiency and spoof-resistance tolerable to the community, this can be done as one time work from a model or empirical data (see for example Figure \ref{threshold-P}). Solutions based only on Hamming distance (not requiring these parameters) are also shown to be viable---although these are typically of lower efficiency and less resilient to spoofing. A compromise in rule complexity combined with one-time parameter tuning, would be essential in a best deployed solution.
  
\subsection{Redefining Chainwork}

Before introducing any changes to the blockchain, let us first analyze how a standard blockchain algorithm like Bitcoin would be affected by probabilistic hash generation and validation. We assume that the miner proposes their hash without considering their confidence value ${\cal P}_{\rm miner}$, and the validator accepts the hash if all bits match. In the limit of rare bit-flipping, at least one bit may deviate from its theoretical (deterministic) value. This introduces an additional mechanism for chain splitting, which we call ``validation fork". With small probabilities, two key events can happen:
(1) False positive by the miner: A miner believes they have found a valid nonce and broadcasts a block that is rejected by all stakeholders. This occurs with probability $1{-}{\cal P}_{\rm miner}$.
(2) False negative by an isolated validator: An individual stakeholder fails to reproduce a valid block accepted by all other stakeholders, with probability $1{-}{\cal P}_{\rm validator}$.
In Bitcoin, the strongest chain is defined as the sequence of blocks consistent with the most valid work. Any path containing a single invalid block cannot be part of the strongest chain from the stakeholder's perspective. Under this consideration, we can examine the consequences.

Scenario (1) would lead to a soft validation fork. The miner's proposed branch becomes obsolete as a new branch mined by the dominant computing power quickly replaces it. The isolated miner rejoins the strongest chain with high probability after two block broadcasts. This soft fork is similar to those caused by network desynchronization in Bitcoin and is not problematic other than it results in some unrewarded work that must be accounted for in mining rewards.

Scenario (2), on the other hand, can cause a hard validation fork. Since the stakeholder cannot join a chain with invalid transactions, they become permanently disconnected from the main chain. A stakeholder who fails to validate a block is left waiting for a new fork that may emerge slowly (or not at all), and regardless is unlikely to be considered the strongest chain by others. If this issue persists, stakeholders will eventually split into hard forks, causing the blockchain to stop functioning. To prevent this, the definition of the strongest chain must allow for the inclusion of unverified blocks.

We can modify the definition of the strongest chain to resolve the hard-validation fork problem. The strongest chain remains defined as the path with the greatest cumulative work, but we allow the block tree to contain invalid blocks by attributing negative work to them. The simplest approach is to introduce equal but negative work for invalid blocks: 
\be
{\cal W}_n = 
 \left\{
\begin{array}{cc} \label{softreject}
 2^{{\cal N}_{\rm zeros}}, & {\rm valid \ block} \\
 - 2^{{\cal N}_{\rm zeros}}, &{\rm invalid \ block}
\end{array} \right. .
\ee
If the work remains constant, then one can assign $\pm 1$ to valid/invalid blocks, we call this basic +/- Chainwork. In the case of a tie, the first path evaluated is preferred, following the same rule as in Bitcoin. This adjustment effectively resolves the issue by turning a hard validation fork to a soft validation fork that is resolved subject to a delay. 
If a user falls behind by two blocks, the majority compute power of miners will produce a third block, causing the strongest chain to advance by three blocks in agreement with the majority.

A more sophisticated way to modify Bitcoin's original Chainwork calculation is to incorporate validator's confidence in the hashes. This can be easily done by reweighting the work (per block) in proportion to confidence:
\be
{\cal W}_n = 2^{{\cal N}_{\rm zeros}} {\cal P}_{\rm validator} = 2^{{\cal N}_{\rm zeros} - {\cal N}_{\rm validator}}.
\ee
This is equivalent to subtracting the number of uncertain bits from the required number of zeros when calculating the work. Note than in the limit of ${\cal P}_{\rm validator} \to 1$ or ${\cal N}_{\rm validator} \to 0$, ${\cal W}_n$ becomes identical to that defined by the standard Bitcoin algorithm. Conversely, if the validator has no confidence in the hash, then 
${\cal W}_n = 0$. Assigning zero work to invalid blocks can open the network to attacks such as spamming with incomplete work. To prevent this we can again introduce a threshold ${\cal N}_{\max}$ for the allowed discrepancy. The validator may reject the block if ${\cal N}_{\rm validator} > {\cal N}_{\max}$ or assign a negative work to invalid blocks similar to \eqref{softreject}:
\be \label{NmaxReject} 
{\cal W}_n = 
 \left\{
\begin{array}{cc}
 2^{{\cal N}_{\rm zeros} - {\cal N}_{\rm validator}}, & \ {\cal N}_{\rm validator} \leq {\cal N}_{\max} \\
 - 2^{{\cal N}_{\rm zeros}}, & \ {\cal N}_{\rm validator} > {\cal N}_{\max}
\end{array} \right. .
\ee
We call this confidence-based Chainwork.
In both cases, no block with zero work would be added to the blockchain. Modifications to the definition of work impact blockchain efficiency, transaction delays and susceptibility to attacks as discussed in discussed in Appendix \ref{attacks}.

\section{Demonstrated Implementation}
\label{ExperimentalImplementationMainSection}
Realization of a quantum blockchain requires access to a quantum computer capable of solving complex problems where quantum advantage has been demonstrated or is potentially achievable. The generated distribution must exhibit sufficient structure to allow observables with nontrivial expectation values that can be measured reproducibly across multiple QPUs. Therefore, techniques such as boson sampling or random circuit sampling which cannot be formulated as efficient decision problems, are not suitable for this application (see Appendix \ref{RCS}). Coarse-grained boson sampling has been proposed as a means to implement blockchains~\cite{shi_quantum_2022, singh_proof--work_2025}, but to our knowledge there is no demonstrated implementation of a blockchain using quantum computers (see Appendix~\ref{CGBS}). Currently, Ref.~\cite{supremacy} is the only experimental demonstration of beyond-classical computation that meets the requirement of efficient cross-validation. Therefore, we base our demonstration on the same protocol. Since Ref.~\cite{supremacy} has already established agreement with quantum mechanics and the infeasibility of classical simulation, our primary focus here is to demonstrate that multiple QPUs can generate and validate hashes, ensuring the blockchain operates reliably with high efficiency and small transaction delays. 

\begin{figure}
\includegraphics[width=\linewidth]{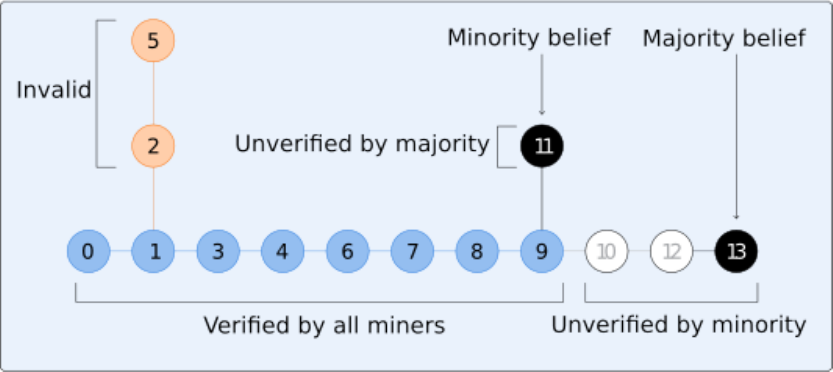}
\caption{\label{typical-blockchain-14nodes} Operation of an example blockchain with $50$ miners using four generally-available Advantage and Advantage2 QPUs. The mining difficulty was set to $\mathcal{N}_{\rm zeros}=32$ with basic +/-1 Chainwork defined in \eqref{softreject}.  
Orange blocks show soft (resolved) forks, these blocks are included in no strongest chains (have been rejected by all miners). Blue blocks are immutable, included in all strongest chains (agreed upon by all miners). Gray and black blocks are in contention: common to some (but not all) strongest chains. Only black blocks are candidates for maximum Chainwork (are being mined) and have the potential for further branching. 
Blocks are labelled by the order of broadcast with the genesis block on the left and most recently mined block on the right. Either of the paths terminating in black nodes might be extended, but the orange path is no longer a candidate for the strongest chain. Blocks 10, 11, 12 and 13 contain delayed transactions, that require further mining (fork resolution) in order to be confirmed. A majority of miners believe node 13 has maximum Chainwork, and in this demonstration all miners consolidate on this branch as the chain develops (see Figure \ref{typical-blockchain}).}
\end{figure}

\begin{figure}
\includegraphics[width=\linewidth]{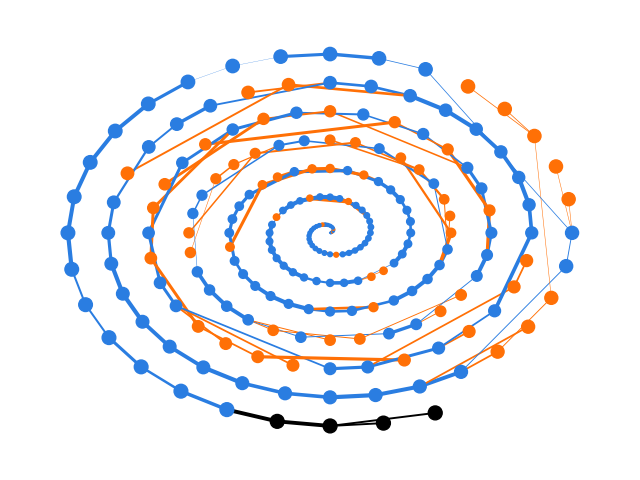}
\caption{\label{typical-blockchain} The same blockchain as described in Figure \ref{typical-blockchain-14nodes} after 219 block broadcasts with matching color scheme. The data was collected over a two-day period. The mining process is accelerated without impact on statistical outcomes (further details and results are contained in Appendix \ref{ExperimentImplementationSection}). The thickness of lines is proportional to the number of miners transferring to the block at the time of its broadcast. The genesis block is placed centrally with blocks placed on a spiral in the order of broadcast. Approximately 70\% of block broadcasts are immutable (blue).}
\end{figure}

\begin{figure}
\includegraphics[width=\linewidth]{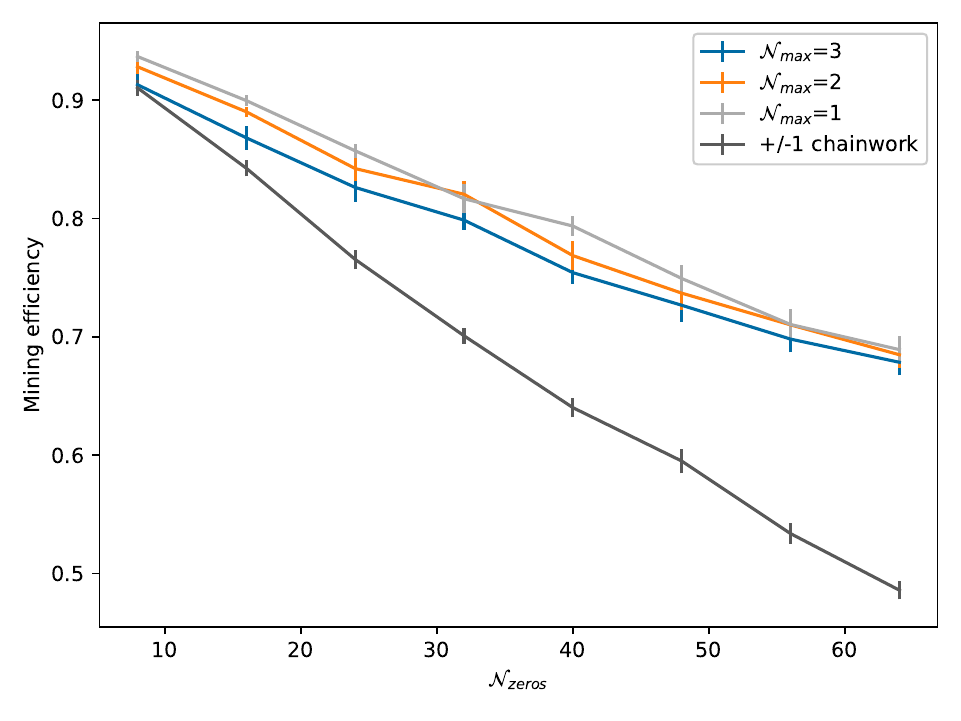}
\caption{\label{threshold-P} Mining efficiency, i.e., the fraction of blocks joining the strongest chain. The Chainwork uses either the simple +/-1 weighting, or the confidence-based weighting with $\mathcal{N}_{\rm max}=1,$ $2$ or $3$ with a constant $\delta W_{\alpha}$ extracted from demonstration data. Confidence-weighted strongest chains are more efficient. The mining process is accelerated without impact on statistical outcomes as described in Appendix ~\ref{ResamplingWitnessesSection}. Results are obtained with resampling from demonstration-parameterized witness distributions to accelerate analysis, each point an average on 16 chains of length 512-1024. Resampling statistics are in agreement with full-demonstration efficiencies reported in Figure \ref{typical-blockchain} and appendices. Further details and statistics are contained in Appendix \ref{ExperimentImplementationSection}.}
\end{figure}

Blockchains, whether our proposal or classical, operate according to rules agreed by a community. Ideally the rule set should be small and persistent, so that rule changes that might cause forking of the user base are infrequent. As with all practical blockchains, rules should be calibrated and agreed based on the capabilities of near-term participants, and revisited periodically by the community. We test several rule sets building upon the well-established Bitcoin rule set, but excluding some essential components: such as periodic updates of problem hardness, modelling of network desynchronization and participation of fraudulent actors. Relative to Bitcoin there are additional parameters; most importantly the choice of an ensemble of unitary evolutions. Our proof of concept demonstrates the feasibility of high efficiency and low delay blockchains in this context, but leaves open the possibility to further optimize protocol rules as a function of the deployed computing environment. We note that significant practical enhancements in performance are achievable with little overhead by employing the mitigations outlined in Appendix \ref{attacks}. 

The demonstrations are conducted on four D-Wave Advantage and Advantage2 QPUs. Each QPU comprises a network of coupled qubits that undergo continuous evolution governed by a time-dependent Hamiltonian 
\begin{equation}\label{HS}
  H(s) = -\Gamma(s)\sum_i \sigma_i^x  + {\cal J}(s)  H_P\;,
\end{equation}
where the problem Hamiltonian is
\be \label{HP}
H_P = \sum_i h_i \sigma_i^z + \sum_{\langle i,j\rangle} J_{ij} \sigma_i^z\sigma_j^z\;,
\ee
and $s=t/t_a$, with $t$ being time and $t_a$ being the annealing time. The energy scales $\Gamma(s)$ and ${\cal J}(s)$ evolve with time in such a way that $\Gamma(0) \gg {\cal J}(0)$ and $\Gamma(1) \ll {\cal J}(1)$. These energy scales are predetermined during calibration and typically vary between different QPUs. The dimensionless parameters $h_i$ and $J_{ij}$ are programmable. The inter-qubit connectivity provided by Advantage and Advantage2 are different, but they are compatible over the subgraphs used for the study.

The quantum problem we solve for hash generation is the coherent quenching of spin glasses through quantum phase transitions. We embed cubic lattices of size $4 {\times} 4 {\times} 4$ dimers (128 qubits) on each of four different QPUs in Canada and the United States; supplementary results for $18 {\times} 18$ dimerized Biclique models (72 qubits) are limited to Appendix \ref{ExperimentImplementationSection}. The critical dynamics ensure that by adjusting a single parameter (annealing time) different QPUs can produce consistent results despite variations in their energy scales~\cite{supremacy}. All qubit biases are set to zero, $h_i = 0$, and couplers are selected as a function of the input message $J_{ij} = \Theta(\mathcal M)$. This is done in two steps: First the message goes through a classical cryptographic hash, e.g., SHA-256 in our case. Then the classical hash is used as a seed in a pseudo-random number generator that generates the unitary-evolution parameters ($J$ only, in our proof of concept). The classical hash is also used as the cryptographic identifier of the block in the block header of the next block. In our proposal based on the Bitcoin block (Figure \ref{block-structure}) the quantum hash is only used for PoQ. This way, we can choose ${\cal N_H} = {\cal N}_{\rm zeros}$ because nonzero bits in the quantum hash have no significance (see Appendix \ref{SectionEnhancedBlockStructure} for a mitigation of this constraint). Using a spin-glass Hamiltonian for quantum evolution ensures that classical collision resistance carries over to the quantum hash. It is extremely unlikely that two spin-glass Hamiltonians with entirely different randomly chosen parameters collide in the resulting high-dimensional distribution. Furthermore, the projection matrix $G(\mathcal M)$ is also randomly generated based on the message $\mathcal M$, making collisions even more improbable.

The feature vector is constructed to include only pairwise nearest-neighbor correlations: $V = [...,\langle \sigma^z_i \sigma^z_j \rangle,... ]^T$. As demonstrated in \cite{supremacy}, reproducing $V$ within the accuracy of QPUs is computationally infeasible for classical computers. The witness vector $W$ is calculated using random projection by a matrix $G$ with matrix elements selected from a normal distribution $N(0,I)$; elements are pseudo-random functions of ${\mathcal M}$. Thresholds ($W_0$) for the witnesses are set as zero. As a miner, we run the quantum algorithm multiple times, each time with a different nonce until we find a hash with $\mathcal{N}_{\rm zeros}$ bits equal to zero (more specifically for confidence-based Chainwork, a hash satisfying the confidence threshold). The miner then adds the block to the blockchain with the nonce that generated ${\cal H} = 0$. The validator then goes through the same calculations, i.e., generates the classical hash and thereby the QPU parameters and the quantum hash and checks if all bits are zeros. 

Fig.~\ref{typical-blockchain-14nodes} presents the initial-stage operation of a blockchain using four generally-available D-Wave Advantage and Advantage2 QPUs, involving 50 miners with $\mathcal{N}_{\rm zeros} = 32$. The visualization captures the dynamic evolution of the blockchain, highlighting block statuses and mining activity. Blocks on the (majority) strongest chain are arranged chronologically from left to right, with the genesis block on the far left and the most recently mined block on the far right. Different block colors indicate their status within the consensus process: Blue blocks (immutable) are included in all strongest chains and are universally agreed upon. Orange blocks (soft forks) have been rejected by all miners and are not included in any of the strongest chains. Gray and black blocks appear in some but not all strongest chains, reflecting temporary disagreement among miners.
Only black blocks are actively being mined, meaning they have the potential for further branching. This illustrates how consensus emerges dynamically as blocks are proposed and either accepted or rejected by the network.  

In Fig.~\ref{typical-blockchain}, the blockchain is shown after 219 block broadcasts, with the genesis block repositioned centrally for clarity. By this stage, approximately 70\% of proposed blocks have become immutable (blue), aligning well with theoretical expectations based on the bootstrapping methodology (see Appendix \ref{ResamplingWitnessesSection}). The remaining blocks continue to be evaluated, with only those on the strongest chains persisting in the blockchain ledger.
This simulation demonstrates the robustness of the blockchain framework, showcasing its ability to maintain stability and achieve miner consensus despite the probabilistic nature of quantum computation. We are able to use statistical properties of the Chainwork described in Section \ref{MiningRateSection} to eliminate failed mining events, and create the blockchain with only $50\times 219$ total QPU programmings. Blockchains with more miners, and over $1000$ block broadcasts, are shown in the appendices.

Fig.~\ref{threshold-P} illustrates the mining efficiency across multiple blockchain operations. Mining efficiency is defined as the fraction of block broadcasts, excluding the genesis block, that successfully enter the strongest chain in the limit of many miners. For each curve the data is taken from 16 chains created with bootstrapping of witness distributions, each having between 512 and 1024 block broadcasts. The bootstrapping method is in agreement with results without bootstrapping reported in other figures. In Appendix \ref{StatisticsSection} we describe mining efficiency and transaction delay in greater detail, presenting additional data. The analysis compares different Chainwork definitions and post-processing methods. For the basic $\pm 1$ weighting method, which assigns equal weights to all block broadcasts, mining efficiency drops to close to 50\% as $\mathcal{N}_{\rm zeros} \to 64$.
The confidence-based weighting technique performs better, as expected, approaching 75\% efficiency for $\mathcal{N}_{\rm zeros}=64$. 
Confidence-based Chainwork allows higher efficiency than basic +/-1 Chainwork, also when accounting for small differences in the mining rate as outlined in Appendix \ref{MiningRateSection}.
To calculate confidence values we used a constant $\delta W_\alpha$ extracted from demonstration data, as detailed in Appendix \ref{ExperimentImplementationSection}.

We would emphasize that per nonce there is a unitary evolution (QPU programming), which in our proof of concept requires 1 second of QPU access time on a 12.5kW quantum computer~\cite{powerconsumptionpaper}. Per nonce there is also a need to perform one classical cryptographic hash for generation of unitary parameters, amongst other low-cost classical computations (generation of unitary evolution parameters, summation of read-out states for statistical estimation, and locality-sensitive hashing [projection of statistics onto witnesses by random vectors]). If the cryptographic hash is calculated on a modern CPU, for example SHA-256, this requires of the order of several hundred clock cycles, or approximately 100 ns\cite{costofclassicalhashing}. Even if one were to consider shorter (but still practically possible) quantum demonstrations, the classical costs per nonce attempt (including the classical hash) are negligible by comparison with operation of the quantum computer in terms of time, energy and dollars.

In this context we can also highlight significant advantages in terms of energy consumption by comparison with classical proof-of-work blockchains operating at equivalent block value. A quantum computer can cost on the order of 10 million US dollars and operate on the order of years\cite{costofquantumcomputers}. The cost is therefore of the order of thousands of USD per hour, by contrast with the energy cost of 12.5kW~\cite{powerconsumptionpaper}, which is on the order of 1 USD per hour. The dollar value of computer power brought to bear in mining is expected to scale with the value of the block in both classical and quantum blockchains, if sufficient supply can be achieved.  Electricity costs are estimated to comprise 90\% to 95\% of Bitcoin's total mining expenses \cite{BitcoinEnergyRatio}. In quantum computing the relative energy cost would be of the order of $O(1/1000)$.

In combination with a reference Python implementation, witness distributions for the four QPUs used in this study for all resampling analyses are available in Zenodo at \cite{Zenodo}. The reference implementation allows construction of blockchains with either generally-available QPUs, or reproduction of the bootstrapping methodology using the data collected for this study.

\section{Conclusions}

We have proposed and successfully tested a blockchain architecture in which proof of quantum work (PoQ) is performed on quantum computers. Our approach introduces multiple quantum-based hash generation methods, with varying levels of complexity depending on the capabilities of the quantum hardware. To address the inherent probabilistic nature of quantum hash generation and validation, an unavoidable property of any quantum hash algorithm, we developed techniques to ensure blockchain stability. We tested our proposed blockchain on four D-Wave annealing QPUs, from two processor generations located in the United States and Canada. We demonstrated that despite architectural and parametric variations, these QPUs can be programmed to generate and validate each other's hashes in a way that allowed stable blockchains to persist over thousands of broadcast blocks and hundreds of thousands of hash validations.

For this demonstration, we used the critical dynamics of 3D and biclique spin glass at a fixed size as the computational problem for PoQ. This was motivated by its role in the first demonstration of beyond-classical computation in a real-world problem~\cite{supremacy} that is not based on boson or random-circuit sampling. However, this choice does not constrain the quantum problem that can be used for PoQ. By choosing a sufficiently complex ensemble of unitary evolutions, in combination with sufficient sampling and post-processing, statistical estimation to high precision by classical computers is made infeasible. Strategies, such as randomizing problem topology, dimensionality, qubit count, and annealing time, can prevent classical learning algorithms from predicting expectation values based on large datasets \cite{GooglePowerOfData}. Additionally, hash algorithms leveraging entanglement witnesses and shadow tomography can further enhance the quantum nature of the hashing process. We note that while beyond-classical computation is desirable for PoQ, it may not be strictly necessary as long as classical competition remains prohibitively expensive, making quantum approaches the only practicable ones. In fact, one may deliberately use smaller problem instances where the ground truth can be obtained --- albeit with significant computational effort --- to benchmark and calibrate QPUs.

A quantum blockchain could transform existing blockchain technology by significantly mitigating the substantial electricity demands of classical proof-of-work (PoW) mining.
In our PoQ approach, energy consumption amounts to a small fraction of quantum computation cost (which is the dominant cost in the PoQ approach). For example, energy expenses account for just 0.1\% of the total computation cost on D-Wave quantum computers~\cite{powerconsumptionpaper}. Replacing classical PoW with quantum PoQ could therefore reduce energy costs by a factor of 1,000. This ratio may shift as quantum computation becomes more cost-efficient over time, but substantial energy savings are expected to persist. Transition to PoQ would also relocate mining from regions with low electricity costs to countries with advanced quantum computing infrastructure. Quantum computation has the potential to undermine the security of some proof-of-work blockchains, unstructured database search (Grover's algorithm) being the best-known example~\cite{10.1145/237814.237866}. Unitary evolution in the beyond-classical regime is at the core of our algorithm; Grover's algorithm requires an efficient oracle to determine membership, which cannot be constructed in our case, and thus a layer of security is provided. Beyond demonstrating the first practical use of beyond-classical computation in a critical domain, this work proves the feasibility of quantum computing in real-world tasks and suggests broader high-impact applications with today's technology.

\section*{Data availability}
A limited set of experimental data is available in the Zenodo online repository, as well as a reference software implementation of methods\cite{Zenodo}.

\section*{Acknowledgment}

We thank Brian Barch, Vladimir Vargas Calderon, Rahul Deshpande,  Pau Farr\'e, Fiona Hannington, Richard Harris, Emile Hoskinson, Trevor Lanting, Daniel Lidar, Peter Love, Chris Rich, Murray Thom and Gunnar Miller for fruitful discussions and comments on the manuscript.
\cut{
\bibliography{Bibliography} 
\bibliographystyle{ieeetr}
}

\appendix

\section{Cross-Validation Limitations in Boson and Random-Circuit Sampling}
\label{RCS}

To operate within a quantum blockchain, QPUs must cross-validate each other's outcomes~\cite{nikolopoulos_cryptographic_2019}. 
Unstructured distributions yielding states with close to uniform random probability, such as those obtained from random-circuit sampling, are unsuitable for this purpose: the expectation values of all foreseeable k-local operators remain close to zero, making them ineffective in the construction of witnesses. The only observable with a nontrivial expectation is $\bra{\psi}\rho \ket{\psi}$, where $\ket{\psi}$ is a ground-truth wave function. For this reason the estimator commonly used to establish beyond-classical computation is the cross entropy~\cite{XEB}. Given a distribution allowing fair and efficient sampling in the computational basis ($P_{\text{QPU}}$, typically a QPU), and a well-resolved ground-truth distribution in the same basis  ($P_{\text{GT}}$, used for verification), the cross entropy is quantified as
\be
\text{XEB} = 2^n \sum_{x} P_{\text{GT}}(x) P_{\text{QPU}}(x) - 1 \;.\label{XEB}
\ee 
$P_{\text{GT}}$ is typically computed using tensor-network techniques at small scales.
This allows the sum to be evaluated by Monte Carlo sampling (of $P_{\text{QPU}}$, i.e. use of a sample set). 
Calculating $P_{\text{GT}}$ by classical methods becomes intractable at large system sizes. The alternative is then to determine $P_{\text{GT}}$ from a QPU. XEB is small in experimental demonstrations, even using an ideal (tensor network) distribution as $P_{GT}$~\cite{Google2024, Chinese2025}. Nevertheless, we might hope that $P_{GT}$ could be constructed for the purpose of validating quantum work.

Unfortunately, the properties of boson sampling and random-circuit sampled distributions mean that the summation of (\ref{XEB}) cannot be executed in a scalable way, even by Monte-Carlo methods. Since sampled state probabilities are all close to $1/2^N$ (for $N$ spins) establishing $P_{GT}$ with sufficient resolution requires $O(2^N)$ samples. More generally, for any pair of high entropy distributions amenable to fair sampling (but not exact integration), impractically large bias in estimation of distributional distance can be an issue\cite{XEB, supremacy}. Entropy is especially large in boson and random-circuit sampling.

\subsection{Coarse-grained boson sampling blockchains}
\label{CGBS}

A proposal for a proof-of-work blockchain using coarse-grained boson sampling (CGBS) was recently proposed by Singh et al.~\cite{singh_proof--work_2025}. The form of unitary evolution and measurement that is undertaken is defined by boson sampling. In order to bypass the problem of resolving state probabilities, samples are binned; the statistic to be cross validated is then the most-probable bin, which with care and sufficiently (polynomially many) samples may be reproducible given a set of devices of comparable (high) control~\cite{nikolopoulos_cryptographic_2019, anguita_experimental_2025}. Miners' broadcasts are synchronized and staked (a miner pays per broadcast), and results are collected subject to a classical-filtering process (mode binning) into a single ground truth. If miners submit sample sets well-matched to the ground truth they are rewarded in proportion to the number of samples broadcast (work). Useful bounds are derived as a function of sampling error and control (photon and other losses), making clear that further development of these devices is necessary for a demonstration. 

Binning of uncorrelated nearly-equal probability states results in a concentration of probability, so that impractically many samples are required to resolve the differences amongst bins---the problems outlined in the previous section are not bypassed by binning in the case of random-circuit sampling. However, in boson sampling evidence suggests that enough structure exists in the distribution to allow a resolution of the most-probable bin with a number of samples scaling weakly with the problem size~\cite{nikolopoulos_cryptographic_2019}. Mode-binned distributions, and k-local correlations in the modes, allow for classical reproduction. Mode-binning is proposed by Singh et al. as a prefilter, with the final proposal being based on state-binning that is robust to classical approximation. State-binning must be done very carefully to allow efficient non-spoofable cross-validation on near-term devices, for a plurality of possible unitaries (bin permutations). 

The consensus mechanism based on CGBS faces several challenges including an absence of devices operating in the beyond-classical regime, and the need to improve quality of device. The collective ground-truth consensus mechanism proposed is quite different from the Bitcoin-like scheme we demonstrate. Whereas the evaluation of work in our proposal requires that a stakeholder trust their own verifying device, in the CGBS proposal there is a need to trust an aggregated ground truth across many miners built with a requirement for strong network synchronization. Several interesting mechanisms are proposed to mitigate for weakness of the protocol, some similar to those in Section \ref{attacks}. In particular we note that with the combination of a stake and an efficient classical filter, many obvious attacks, based on manipulation of the ground truth, become impracticable. Further development may be require to mitigate for attacks that involve recombining, resubmitting and/or buffering samplesets for enhanced reward (attacks which even when unrewarded may destabilize verification). Note that states might be manipulated in a manner preserving exact or approximate mode populations. There are also weaknesses in our blockchain arising from, for example, the redefinition of Chainwork that do not impact this proposal.

\section{Witnesses Construction Beyond the Computational Basis}
\label{WitnessConstructionBeyondTheComputationalBasis}

\subsection{Multi-Basis Measurement}

In some quantum systems, it is possible to perform measurements in multiple bases. This is feasible, for example, when the system allows single-qubit rotations (whether Clifford or non-Clifford gates) prior to measurement. Both digital and digital-analog QPUs can potentially enable such capabilities. Measuring in multiple bases allows access to off-diagonal elements of the density matrix, $\rho_\mathcal{M}$. Given their exponential growth in number, accessing all off-diagonal elements is both impractical and unnecessary for most applications. Therefore, rather than attempting to recover all elements of the density matrix, one can measure quantities accessible through sampling by performing measurements in multiple bases. Digital-analog methods can be realized in experimental D-Wave quantum annealing processors using multi-color annealing methods for purposes of measuring non-computational basis information\cite{digital-analog}.

The easiest way of using multi-basis measurement is by randomizing the measurement basis. The basis to measure the qubits can be determined by the message through the QPU parameters $\Theta(\mathcal M)$. This way random basis rotations are considered part of the unitary evolution $U_\theta$. While this would make classical simulation very challenging, it is still based on a single basis measurement of the density matrix $\rho_\mathcal{M}$ and the previous concern persists.

A better option is to measure the same $\rho_\mathcal{M}$ in different bases. This allows calculating entanglement witnesses as components of the witness vector. An entanglement witness is a measurable quantity, $W$, such that exceeding a threshold value, 
$W>W_0$, necessarily implies the presence of entanglement while $W<W_0$ does not guarantee the absence of entanglement. We can design a quantum hash such that each $W_\alpha$ in Eq.~\eqref{Halpha} represents an entanglement witness, and each corresponding $W_{0\alpha}$ acts as a boundary, beyond which entanglement is assured.

As an illustrative example, we consider using Bell's inequality violations for pairs of qubits as entanglement witnesses. In an appropriate experimental setup, a loophole-free violation of Bell's inequality indicates that the experimental results cannot be explained by any local hidden-variable theory. However, this is not our objective. Instead, we aim to use the inequality as an entanglement witness. Specifically, we employ an experimentally feasible variation of Bell's inequality, known as the Clauser-Horne-Shimony-Holt (CHSH) inequality.

\subsection{CHSH Inequality}
\label{CHSHInequalitySection}

In this appendix we introduce CHSH inequality and formularize it in a way suitable for our application. Consider two observers $a$ (Alice) and $b$ (Bob), both allowed to perform measurement in two orthogonal bases ``1'' and ``2'' via operators $a_1,a_2,b_1,b_2$. We define the Bell's operator as
\be
B = a_1b_1 + a_1b_2 + a_2b_1 - a_2b_2
\ee
Consider a quantum state represented by a density matrix $\rho$. Measuring the Bell operator would result in an expectation value $\langle  B \rangle_\rho = {\text Tr}[\rho B]$. Bell's theorem states that for any classical local hidden variable theory, we should have
\be
| \langle B \rangle_\rho| \leq 2 \qquad {\text Classical},
\ee
whereas quantum mechanics allows 
\be
| \langle B \rangle_\rho| \leq 2\sqrt{2} \qquad {\text Quantum}. 
\ee
This is commonly known as the Clauser-Horne-Shimony-Holt (CHSH) inequality, an experimentally practical variation of Bell's inequality \cite{CHSH}. Violation happens when $\langle B \rangle_\rho > 2$, which indicates that the experimental results cannot be explained by any local hidden variable theory. It is clear that violation of CHSH inequality is a sufficient but not necessary condition for quantum mechanics. It is also evident that violation strongly depends on the choice of $a$ and $b$ operators as well as the quantum state $\rho$. We therefore define an operator-independent quantity as a maximum over all possible operators:
\be
\langle {\text{CHSH}} \rangle_\rho = \max_{A_\alpha,B_\beta}{\text Tr}[\rho B].
\label{CHSH}
\ee
The CHSH inequality therefore is expressed as $\langle {\text{CHSH}} \rangle_\rho \leq 2$. 

Consider the Bell state
\be
\ket{\psi} = {\ket{\uparrow\uparrow} + \ket{\downarrow\downarrow} \over \sqrt{2}}
\ee
where $\ket{\uparrow}$ and $\ket{\downarrow}$ are eigenstates of $\sigma^z$ with eigenvalues $\pm 1$, respectively. The pure state density matrix is
\be
\rho = {1 \over 2}(\ket{\uparrow\uparrow} + \ket{\downarrow\downarrow})(\bra{\uparrow\uparrow} + \bra{\downarrow\downarrow})
\ee

We define operators
\ba
a_1 = \sigma_a^x, \ a_2 = \sigma_a^z, \  b_{1} = {\sigma_b^x {+} \sigma_b^z \over \sqrt{2}}, \ b_{2} = {\sigma_b^x{-}\sigma_b^z \over \sqrt{2}},
\ea
which means Bob's measurement bases are 45 degrees rotated compared to Alice. The Bell operator therefore becomes
\ba
B = \sqrt{2}(\sigma_a^x \sigma_b^x + \sigma_a^z  \sigma_b^z )
\label{Bell2}
\ea
It is easy to check that $\ket{\psi}$ is an eigenfunction of $B$ with eigenvalue $2\sqrt{2}$. To violate Bell's inequality with  $\langle B \rangle_\rho > 2$, we need to have strong correlations in both $x$ and $z$ directions. Each correlation can maximally be equal to one so the optimal value is $2\sqrt{2}$, which occurs in Bell's state. But weaker correlations can also lead to violation as long as the sum of the two correlators is greater than $\sqrt{2}$.

Classically, if we represent each spin by a 2D vector $\vec S=[\cos \theta, \sin \theta]^T$, we obtain 
\ba
B &=& \sqrt{2}(\cos \theta_a \cos \theta_b + \sin \theta_a \sin \theta_b) \nn
&=& \sqrt{2} \cos (\theta_a - \theta_b)
\ea
Maximum happens when $\theta_a = \theta_b$, which gives $|\langle B \rangle |_\text{max} = \sqrt{2}$. This is less than the maximum possible classically based on conditional probability arguments, i.e., $|\langle B \rangle |_\text{max} = 2$.

It is possible to define CHSH inequalities between pairs of qubits in a multi-qubit system. One can use \eqref{CHSH} with $\rho_{ij}$ representing the reduced density matrix for qubits $i$ and $j$:
\be
\langle {\text{CHSH}} \rangle_{\rho_{ij}} = \max{\text Tr}[\rho_{ij} B],
\label{CHSHrho}
\ee
where the maximum is over all possible measurement bases. While each CHSH inequality can be violated, there is a trade-off relation that needs to be satisfied. For a system of $N$ qubits, we should have \cite{CHSHTradeoff}
\be
\sum_{i<j}^N \langle \text{CHSH} \rangle^2_{\rho_{ij}} \leq 2N(N-1)
\ee
Since the number of pairs is $N(N-1)/2$, the equality is satisfied if $\langle \text{CHSH} \rangle_{\rho_{ij}} = 2$ for all pairs. This means that it is not possible to violate all CHSH inequalities at the same time. In other words, if some of the inequalities are violated, there are always others that are not violated. This is closely related to the monogamy of entanglement, which states that if two subsystems are strongly entangled, they should have less entanglement with other subsystems.

\subsubsection{CHSH based witnesses}

Let us consider a system in which measurements can be performed in two Pauli bases, $x$ and $z$. For any two qubits,  $i$ and $j$, we define a modified Bell operator as (see also Appendix \ref{CHSHInequalitySection}):
\be
B_{ij} = \sqrt{2}(\sigma_i^x \sigma_j^x + \sigma_i^z  \sigma_j^z ).
\ee
Based on a purely statistical argument, it is shown that for any local hidden variable theory,  the expectation value of the Bell operator satisfies
\be \label{CHSHEq}
|\langle B_{ij} \rangle | \leq 2\;,
\ee
which is known as CHSH inequality. When Pauli matrices are replaced by classical spin components, we get an even smaller bound, $|\langle B_{ij} \rangle | \leq \sqrt{2}$ (see Appendix \ref{CHSHInequalitySection}). In contrast, quantum mechanics allows $|\langle B_{ij} \rangle |$ to reach values as large as $2\sqrt{2}$. This is possible when $|\langle \sigma_i^x \sigma_j^x \rangle | = |\langle \sigma_i^z \sigma_j^z \rangle | = 1$, which is classically counterintuitive as it means the spins are oriented both in $x$ and $z$ directions at the same time. Quantum mechanically it is possible because of quantum contextuality, i.e., dependence of the outcome on the measurement setup. It is clear that violation of CHSH inequality is a sufficient but not necessary condition for quantum mechanics, and more explicitly for entanglement between the qubit pair. Therefore, CHSH inequality is an entanglement witness. 

CHSH inequality can also hold in a multi-qubit system and therefore can be used as a basis for hash generation. Let us define $W_\alpha = |\langle B_{ij} \rangle |$ in Eq.~\eqref{Halpha} with a threshold value $W_{0\alpha} = 2$.  Here, $\alpha$ represents a pair of qubits in the multi-qubit system. The hash bit ${\cal H}_\alpha$ is then equal to one if the CHSH inequality for the corresponding pair is violated, and zero otherwise. It has been shown that in a multi-qubits system the CHSH inequality must satisfy a trade-off relation \cite{CHSHTradeoff} (see Appendix \ref{CHSHInequalitySection}):
\be
\sum_{\alpha \in \{ {\rm  all \ pairs} \} } W_\alpha^2 \ \leq 2N_Q(N_Q-1)
\ee
The equal sign happens when for all pairs, $W_\alpha = 2$. It is therefore impossible for all pairs to violate the CHSH inequality simultaneously. In other words, if some of the inequalities are violated, there must be others that are not.

To use the CHSH inequality for hash generation, we must first identify $\mathcal{N}_\mathcal{H}$ pairs of qubits for which the likelihood of violating the CHSH inequality is maximized. Hash generation then reduces to executing a quantum evolution parameterized by $\Theta$ and determining which pairs from the predetermined set of qubit pairs violate the CHSH inequality. Each hash bit is assigned a value of one or zero, depending on whether or not the CHSH inequality is violated for the corresponding qubit pair. No classical model can reliably predict these outcomes, making a quantum computer essential for calculating the hash value. This approach ensures compliance with the seventh requirement mentioned earlier: spoof resistance.

\subsection{Shadow Tomography}

If the QPU can perform multi-qubit unitary operations before measurement, shadow tomography can be utilized for hash generation. First introduced by Aaronson \cite{ShadowTomography}, shadow tomography provides an efficient method for estimating quantum observables using a limited number of measurements. Unlike full quantum state tomography, which requires an exponentially large number of measurements to reconstruct the full density matrix, shadow tomography focuses on extracting specific properties of the state with significantly fewer samples. A more practical variant was later introduced by Huang, Kueng, and Preskill \cite{ClassicalShadow}. In their approach randomized measurements are used to generate compact representations, known as classical shadows ($\hat \rho_i$) of the full density matrix ($\rho$) that can be generated and stored efficiently. The expectation value of any observable $O$ can then be estimated by averaging over all classical shadows:
\be
\langle O \rangle \approx {1\over N} \sum_i {\rm Tr}(O \hat \rho_i)
\ee
where $N$ is the number of measurement rounds. The number of  measurements required to estimate $m$ observables with error $\epsilon$ scales as $\log (m) /\epsilon^2$, making it exponentially more efficient than full state tomography.

To generate a classical shadow, we apply a randomly chosen unitary transformation $U$ from a known ensemble (e.g., Clifford group): $\rho \to U \rho U^\dagger$. 
We then perform measurement in the computational basis to obtain a bitstring outcome $b$. A classical shadow of $\rho$ corresponding to the measurement outcome is calculated as
\be
\hat \rho = M^{-1}(U^\dagger \ket{b} \bra{b} U)
\ee
where $M$ is the shadow tomography measurement channel, defined as:
\be
M(\rho) = \mathbb{E} [U^\dagger \ket{b} \bra{b} U].
\ee
The expectation is over all unitaries $U$ and measured outcomes $b$. The inverse channel $M^{-1} $ depends on the chosen unitary ensemble. For example, for Pauli measurements, it has a simple analytical form (see Ref.~\cite{ClassicalShadow} for detail).

To use shadow tomography for hashing, one can choose ${\cal N_H}$ observables $O_\alpha$ from which witnesses are extracted via shadow tomography:
\be \label{Wa}
W_\alpha = {\rm Tr}(O_\alpha \rho_\mathcal{M}) \approx {1\over N} \sum_i {\rm Tr}(O_\alpha \hat \rho_{\mathcal{M}i}).
\ee
Shadow tomography enables the estimation of expectation values for complex observables $O_\alpha$, which cannot be directly measured or reliably extracted from simple measurement protocols within a feasible number of repetitions. One can, for example, choose $O_\alpha$ such that $W_\alpha$ becomes an entanglement witness. For example, let $O = \ket{\psi}\bra{\psi}$ represent a pure state projection such that 
\be
{\rm Tr}(O_\alpha \rho) > \kappa > {\rm Tr}(O_\alpha \tilde \rho)
\ee
where $\tilde \rho$ is a separable density matrix. By choosing $W_{0\alpha} = \kappa$, the extracted $W_\alpha$ defined in \eqref{Wa}, serves as an entanglement witness in Eq.~\eqref{Halpha}, facilitating hash bit generation.

Hashing based on shadow tomography will likely require a fault-tolerant gate-based quantum computer. However, digital-analog quantum computers with sufficient flexibility may have sufficient accuracy without quantum error correction. The details of the unitary evolution leading to $\rho_\mathcal{M}$ and the random unitaries $U$ applied before measurement would depend on the specific QPU architecture and the available quantum operations.

\section{Weaknesses and Mitigations}
\label{attacks}

Many weaknesses for PoW blockchains such as Bitcoin are known, and many mitigations have been developed~\cite{BitcoinWiki, CB6, CB7, CB8, CB9, singh_proof--work_2025}.
We focus on several attacks that might be attempted by an adversarial agent to gain reward with incomplete work. The reward might be gained deterministically, or more often probabilistically (in expectation). Neither the attacks presented, nor the mitigations, are an exhaustive list. 

An example of an attack for PoW blockchains is the 51\% attack. In this attack an attacker sends Bitcoin to a service provider. After a delay (of a few block generation events) the service provider considers the payment made to be immutable, and renders the service. The attacker then creates a fork in the blockchain by mining a block that precedes the payment block, and omits their payment to the service provider in this new branch. If they are able to make this new branch strong (long) enough all stakeholders will switch to it, at which point the transaction to the service provider is no longer part of the record. The attacker is then able to spend their money again (a double spend). If an attacker controls a significant percentage of compute power it can be worth their time to attempt such an attack to negate a large payment, and if they have control of 51\% of compute power they will succeed with high probability.

The 51\% attack and other blockchain weaknesses apply also to our probabilistic blockchain. To mitigate for the 51\% attack we require well-distributed quantum compute power; since mining power can be split by probabilistic verification, the threshold can be lower than 51\% (but still substantial if mining efficiency is high enough). We focus in this appendix on attacks that are new or qualitatively changed relative to classical or non-probabilistic blockchains. Attacks are different and potentially more powerful owing to several factors including a lack of determinism, bypassing of experimental work (inference given the experimental parameters, i.e. spoofing), and changes in the relative costs of broadcasting, mining, and validation. 

As described in the main text, the particular proof-of-concept demonstration is premised on the notion of honest network users and is susceptible to attacks. In addition to mitigations described in the main text we explore additional procedures in this appendix. We present a set of tools to protect against no-work and partial-work attacks, but optimization of the ensemble of unitary dynamics, post-processing and digitalization process, essential in practice, goes beyond the scope of this publication.

\subsection{No-Work Attacks}

We consider first the possibility that a miner might present a fraudulent block, one in which the nonce is chosen uniformly at randomly without conducting any quantum computation or inference. In Bitcoin a random nonce can be used in a broadcast block. This convinces either all users with probability $2^{-\mathcal{N}_{\rm zeros}}$ or convinces no users. The expected reward from such a strategy is $2^{-\mathcal{N}_{\rm zeros}} R - C_B - C_P$, where $C_B$ is the cost of broadcasting the block, $R$ is the expected block reward and $C_P$ the cost of block preparation. Pursuing this strategy repeatedly the cost of block preparation is negligible (only a bit increment of the nonce is required between proposals). By contrast the expected reward from honest mining is $((R - C_B)2^{-\mathcal{N}_{\rm zeros}} - C_P - C_H)$, where $C_H$ is the cost of hashing (validation). The reward in Bitcoin is large enough compared to the broadcast cost and $C_H$ that mining is incentivized. Simultaneously $C_B > 2^{-\mathcal{N}_{\rm zeros}}R$, so random broadcasts are disincentivized. 

The random nonce strategy is similar to a denial of service attack (DoS) from the perspective of validators. In DoS an attacker broadcasts in large volume to overwhelm the bandwidth or verification capabilities of the network (thereby disrupting or delaying transactions). 

DoS and random nonce mitigations effective in other mature blockchain technologies can be used (see e.g. ~\cite{BitcoinWiki}). The situation is qualitatively the same in our probabilistic blockchain, although the expected reward is smaller as a function of the blockchain efficiency (enhanced branching), and validation may be inhomogeneous with respect to users. Validation is also more expensive in our simplest (main-text) implementation, so fewer malicious packets can be more disruptive. The inhomogeneity of validation is a problem, since mining power may be diluted by such an attack even if the attacker is not ultimately rewarded.

In our application we should consider also important practical changes: the cost of hashing $C_H$ is raised, since quantum measurements will be more expensive than single SHA-256 hashes. Similarly $\mathcal{N}_{\rm zeros}$ may be smaller, since the verification time scale is longer. This makes a no-work strategy more viable. Relative to Bitcoin there is a smaller gap between the honest and dishonest expected rewards. To mitigate for no-work attacks a modified block structure can be effective, as described in Section \ref{SectionEnhancedBlockStructure}.

We note that if miners are not anonymous their behavior can be tracked. A rule can be introduced to penalize or temporarily block users who consistently propose unverifiable (or statistically anomalous) work, as is done for DoS. Note that a transmission path or node might be penalized with limited information on the stakeholder. We will later see that deanonymization is not essential to mitigate work and partial work attacks.

\subsection{Partial-work attacks}
Partial work does not offer attack opportunities in Bitcoin owing to the cryptographic properties of SHA-256. A hypothetical probabilistic hash might fail in an independent and identically distributed manner, providing similar guarantees against partial work.
However, our hash is to be inferred from a known unitary evolution ($\Theta$) and digitalization ($G, W_0$) procedure, which in practical implementations can leave room for inference to bypass at least some quantum work.

Consider the case of a blockchain built upon advantage in the estimation of quench statistics~\cite{supremacy}, used in our proof of concept. 
Since signs on hyperplanes are determined as a (classical) cryptographic function of the block header, preimage resistance is assured: it is not possible to determine a block header from a desired (all zero) bit pattern. However, given a nonce (thence protocol parameters $\Theta, G, W_0$) one can estimate correlations with increasing accuracy by a hierarchy of classical computational methods (of increasing cost). Certain unitary evolutions ($\Theta$) may allow better or worse estimators, combined with certain realizations of $G$ and $W_0$ that allow an accumulation of signal relative to the error. In this way the sign $W_\alpha-W_0$ for some (many) bits might be inferred with non-negligible (or perhaps even high) confidence.

An attack might proceed as follows. For many nonces an attacker can estimate $\{W_\alpha\}$ by classical methods (e.g. by approximate simulation, or machine learning). The results are ranked by the number of (robust) threshold-satisfying bits. The best candidate might be broadcast directly. The reward from this strategy is described as in the no-work scenario, but with an enhanced (classical) work cost and enhanced probability that the bits are verified.  This attack is mitigated by the modified block structure of Appendix \ref{SectionEnhancedBlockStructure}. In particular, spoofing enhanced experimental data is impossible given beyond-classical computation~\cite{supremacy}; we demonstrate methods to cryptographically secure and decouple the computational complexity requirement $\mathcal{N}_{\rm zeros}$ from the length of the quantum hash allowing comparison of higher precision experimental results.

A second way to use classical estimators is as a filter. Rather than broadcast the most promising nonce, the nonces are ranked and subjected to the prescribed quantum experiment. Only if the quantum work validates the nonce is it broadcast. The filter allows a user to bypass experiments that are less likely to meet the Chainwork requirement. In this case, the filter success rate must grow slowly enough with the classical work for quantum experiments to be favored over the filter.

A modified block structure can also mitigate for the filter attack as later described. To mitigate this filter attack without a modified block structure it is necessary that all practical estimators are weak. One simple mechanism by which to weaken classical estimators is described in Appendix \ref{HyperplaneDistributionMitigationsSection}. The quality of estimators as a function of the classical-work investigated is above all a function of the unitary-evolution ensemble (distribution of $\Theta$); it is important to choose a unitary evolution with a maximum gap between classical and quantum estimators.

\subsubsection{Partial Quantum Work Attack}
\label{pqwa}
The basic +/-1 and confidence-based chainwork definitions allow for near misses with respect to complete or incomplete work. A participant might have some confidence that their result is close to the threshold, and reasonably  expect that although their validation fails their nonce might be validated by some significant fraction of stakeholders.

Since a finite fraction of stakeholders (thence mining power) can validate a near-threshold block, there is finite probability of such a block being incorporated in the strongest chain subject to a delay; motivating a miner to broadcast such a block. A careful balancing of block penalties and rewards is necessary to ensure quantum-partial work is not a profitable strategy, achievable by control of transaction and block rewards and enhanced block structure.

The quantum computation also allows room for efficiencies, which should be understood in a full application. To minimize QPU access time (costs) a miner should optimize parameters such as number of samples and programming time. They could then more efficiently search, or filter, nonces. A miner is incentivized to prefilter nonces with cheap experiments, and then seek higher confidence (via enhanced sampling) only on candidates for broadcast. In our proof-of-concept demonstrations by contrast we fix QPU access time to 1 second in mining and validation.

\subsection{Multi-Block Attack}
The multi-block attack can be considered a generalization of the 51\% attack. This involves an attacker proposing a no-work (or partial work) block and seeking to bury it under full-work blocks. By burying no-work blocks with full work blocks they may receive a reward enhanced in proportion to the block ratio.

Our definitions of Chainwork mean that for any path to be viable as a strongest chain there must be strictly more verified blocks than unverified blocks. Owing to this, the criteria necessary for this attack to succeed are similar to the standard 51\% attack. In effect an attacker must be able to grow the chain more quickly than alternative non-fraudulent candidates, in order to undo the Chainwork damage of their fraudulent block(s). Although a user might time an attack to coincide with a moment when compute power is fragmented, this still requires control of a significant fraction of community compute power assuming practical chain efficiency (concentrated honest mining power, up to small fluctuations).

\subsection{Enhanced Block Structure Mitigations}
\label{SectionEnhancedBlockStructure}
\begin{figure}
\includegraphics[width=\linewidth]{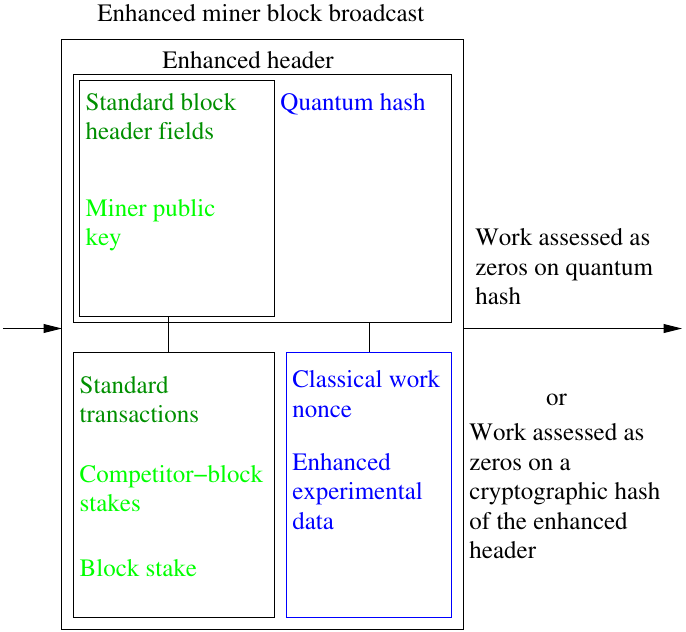}
\caption{\label{block-structure2} The standard block (details in Figure \ref{block-structure}) can be broadcast with additional information or costs to mitigate for weaknesses. Some or none of these enhancements may be useful, depending on the strength of the quantum hash. Dark green and light green are elements that are known prior to the quantum-hash estimation, light being new elements discussed in this appendix. Additional elements in blue can be broadcast and signed by the miner. Since the public key of the miner is protected by completion of the work, it is not possible for anyone but the miner to manipulate bits of the enhanced broadcast without repeating the proof of work.  Completion of supplementary classical work, or inclusion of transactions costs can raise the cost of tampering with this information, and/or the costs associated with fraudulent block broadcasts. One may consider the work requirement with respect to the quantum hash, or a cryptographic hash of the enhanced header. In the latter case, this allows a cheap classical check of block validity, before a second more expensive quantum computation stage must be undertaken.}
\end{figure}

An enhanced block structure shown in Fig.~\ref{block-structure2} can mitigate against several of the attacks discussed. The new block consists of fields known prior to quantum-hash estimation, which impact the parameterization of the quantum-hash estimator, and post-quantum-hash fields, for which manipulation is prevented by the miner's cryptographic signature.

\subsubsection{Broadcasting with Additional Classical Work}
Block broadcasts that are verified with low probability pay a broadcasting cost. In order to raise the cost of broadcasting it can be required that a user completes a classical proof of work. The work should raise costs, yet be negligible compared to the quantum work. If the classical work is sufficiently high, it ceases to become profitable to risk approximations in the quantum part. In the case where the classical work becomes comparable to the quantum work, we have a hybrid system. In an immature technology it may be valuable to pursue such a hybrid approach and gradually reduce the proportion of classical work as confidence is gained in the quantum operation. Quantum proof of work could of course also be hybridized with proof of stake as an alternative mitigation.

\subsubsection{Broadcasting with a Stake}
The cost of broadcasting is also raised if a user is required to stake (bid on) their proposal. The stake can be a finite fraction of the block reward, signed from the miner's current wallet to a burn wallet. Stakes for failed branches are paid to a burn-wallet. In this way, any user is limited in the number of weak proposals they are able to make, and risk significant loses from doing so. Payment to a burn wallet is a standard component in proof-of-stake blockchains such as Etherium, but note that the stake here does not require any community involvement. This is not a proof-of-stake consensus mechanism, which requires a community-wide mechanism. Stake payments from blocks in branches that were ultimately rejected by consensus can later be collected by miners extending the strongest chain, so that they become part of the permanent ledger record. Thus even when a block is rejected, its stake must still be paid; this disincentivizes the broadcast of poor-quality blocks.

\subsubsection{Enhanced Experimental Data}
\label{sec:eed}
We can also raise the difficulty to pass validation by requiring the publication of enhanced experimental data (EED). Quantum work is established based on the leading $\mathcal{N}_{\rm zeros}$ bit description of the quantum hash; by contrast the high-dimensional full experimental data may be subject to more-robust statistical validation tests that detect bad proposals~\cite{supremacy}. Assuming we choose our unitary evolution ensemble so that a plurality of quantum unitary evolutions are not simulatable (to accuracy comparable to that of the quantum-computing participants) classical computing can be eliminated with high probability by choice of a distance threshold (in statistic space, per \cite{supremacy}). EED can be signed by the miner's private key matching the public key that is protected by work. Only the miner can manipulate the experimental data without redoing the work, and is incentivized to demonstrate the best (true) data.

\subsubsection{Enhanced-block Hash for Work Mitigation}
In place of evaluating $N_{zeros}$ on the quantum hash, we may append a quantum hash to the pre-experimental block header and require that a classical cryptographic hash of this object meet the $N_{zeros}$ requirement. This hash can also be used as the block identity, in place of the hash based only on the pre-experimental header components.

This scheme allows for a cheap classical filter that quickly detects manipulation: Before conducting a quantum experiment, or validating other block fields, we can hash the enhanced block header to check it meets the proof-of-work threshold ($N_{zeros}$). This is by contrast with the alternative scheme in which a quantum experiment is required to check the work. Note that this test allows for false positives: it is necessary to check the experimental data via a statistical test to ensure a capable quantum computer actually ran the experiment.

The quantum hash must be consistent with the extended experimental data, must be long enough that a verifier has confidence the broadcaster undertook the valid quantum experiment, yet short enough that miners cannot confidently manipulate the experimental data to attempt many proofs of work from a single quantum experiment (per partial quantum work section \ref{pqwa}). Note that the length of the quantum hash needn't be determined by ${\mathcal N}_{zeros}$, as is the case in our simplest implementation. 

\subsection{Hyperplane Distribution Mitigations}
\label{HyperplaneDistributionMitigationsSection}

\begin{figure}
\includegraphics[width= \linewidth]{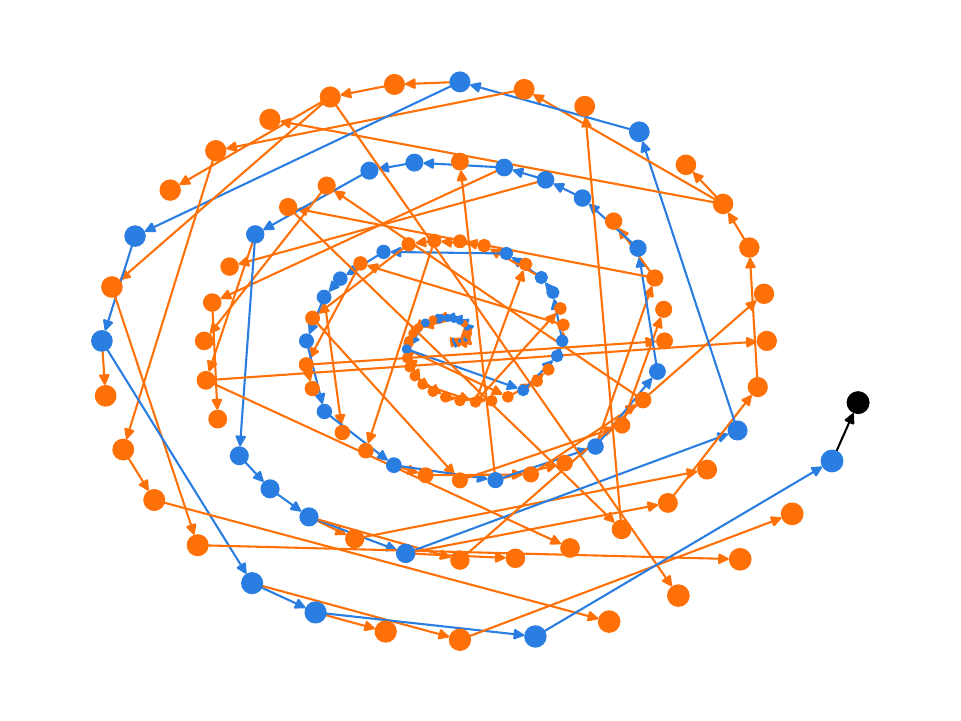}
\caption{\label{fig:confidenceworkPlusJorthog} Operation of an example blockchain of 128 block broadcasts generated in the limit of many miners using projection orthogonal to $J$ with $\mathcal{N}_{zeros}=32$, and confidence-based Chainwork with $\mathcal{N}_{\rm zeros}=1$. The most recent block mined is marked in black and the strongest chain in blue. Averaging efficiency across 5 independent chains yields an efficiency estimate of 0.34 +/- 0.03 in agreement with the bootstrapped estimates of Figure \ref{threshold-P2}.}
\end{figure}

Suppose a per-nonce classical work less than the quantum experimental cost ($C_{H}$) allows for a witness estimator ${\hat W}$. One mitigation is to modify the distribution of thresholds or hyperplanes (to choose $W_0\sim{\hat W}$), or to choose hyperplanes orthogonal to ${\hat W}$. In either case the classical approximation is rendered uninformative on the signs (bits) expected from the quantum experiment. If the uncertainty in the quantum experiment is relatively small, we might still have sufficient accuracy in quantum experiments for robust validation. A visualization of this process is shown in Fig.~\ref{hyperplane projection}.

In the case of our quench unitary dynamics, we know that nearest-neighbor correlation are correlated with $-J$: the approximation $C_{ij} \propto -J_{ij}$ describes accurately the weak coupling (fast anneal) limit. There are significant deviations from this estimate, but any hyperplane that is well correlated with $J$ can accumulate a reliable signal from weak estimators incorporating this approximation. To mitigate for this we might choose our hyperplanes to be orthogonal to $J$. Choosing hyperplanes (or thresholds $W_{0\alpha}$) to decrease reliability of classical estimators will typically enhance the uncertainty in quantum validation, thus reducing the efficiency and increasing delays in a blockchain. This effect is shown in Fig.~\ref{threshold-P2} as a function of $\mathcal{N}_{zeros}$. In Fig.~\ref{fig:confidenceworkPlusJorthog} we show an example of a chain resistant to simple classical inference based on the weak coupling limit approximation. 

\section{Experimental Implementation Details}

\label{ExperimentImplementationSection}
\begin{figure}
\includegraphics[width= \linewidth]{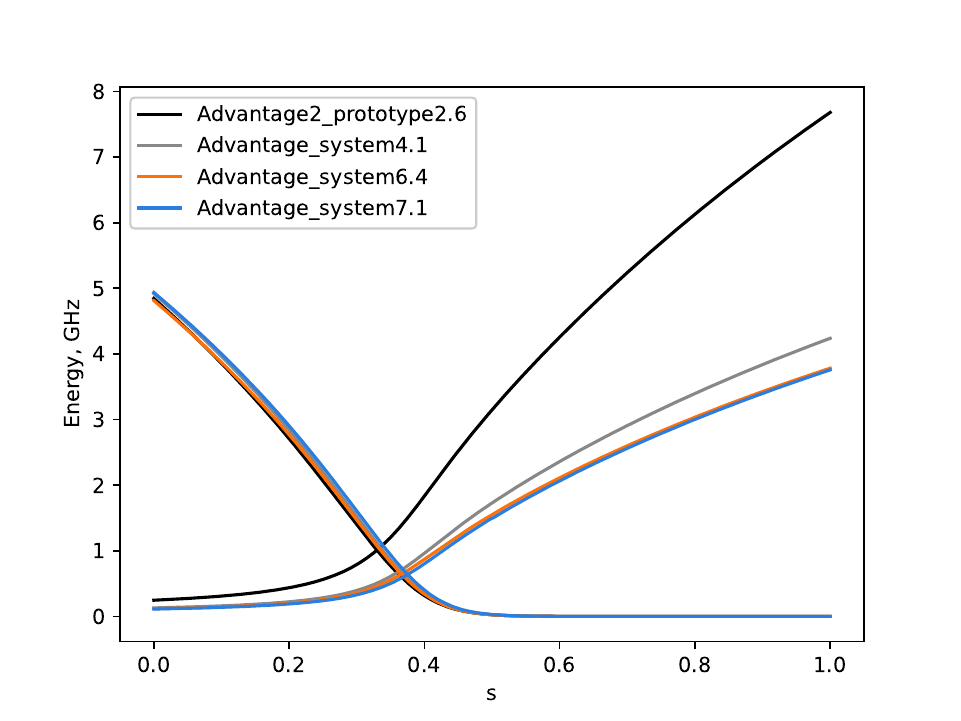}
\caption{\label{schedules} Annealing schedule energy scales of the general-access QPUs used in experiments. $\Gamma(s)$ decreases with anneal progression $s$, ${\cal J}(s)$ increases, to define the time-dependent Hamiltonian \eqref{HS}. Transverse fields are comparable on all devices, but the problem energy scale is larger in Advantage2.  Advantage QPUs can be programmed with (approximately) doubling the anneal duration to simulate an Advantage2 QPU full-energy scale anneal as shown in Figure \ref{collapse}~\cite{supremacy}.}
\end{figure}

\begin{figure}
\includegraphics[width= \linewidth]{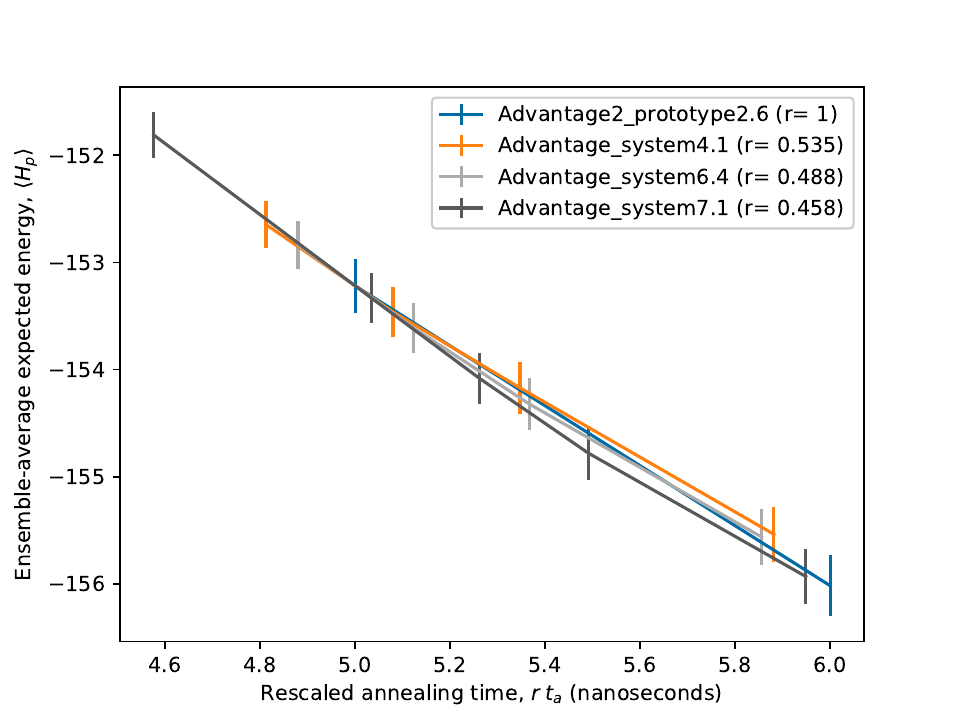}
\caption{\label{collapse} For Advantage systems to emulate Advantage2 annealing processes, where each operates at the maximum programmable energy scales, it is necessary to anneal for longer. These factors can be obtained by a collapse of statistics such as ensemble-average energy, here using averages on 25 random disorder realizations. Factors shown that allow a matching of energy also allow near-optimal cross-validation rates, which is consistent with the unitary dynamics being controlled by critical phenomena, and thereby weakly dependent on the detailed shape of the schedule~\cite{supremacy}.}
\end{figure}

\begin{figure}
\includegraphics[width= \linewidth]{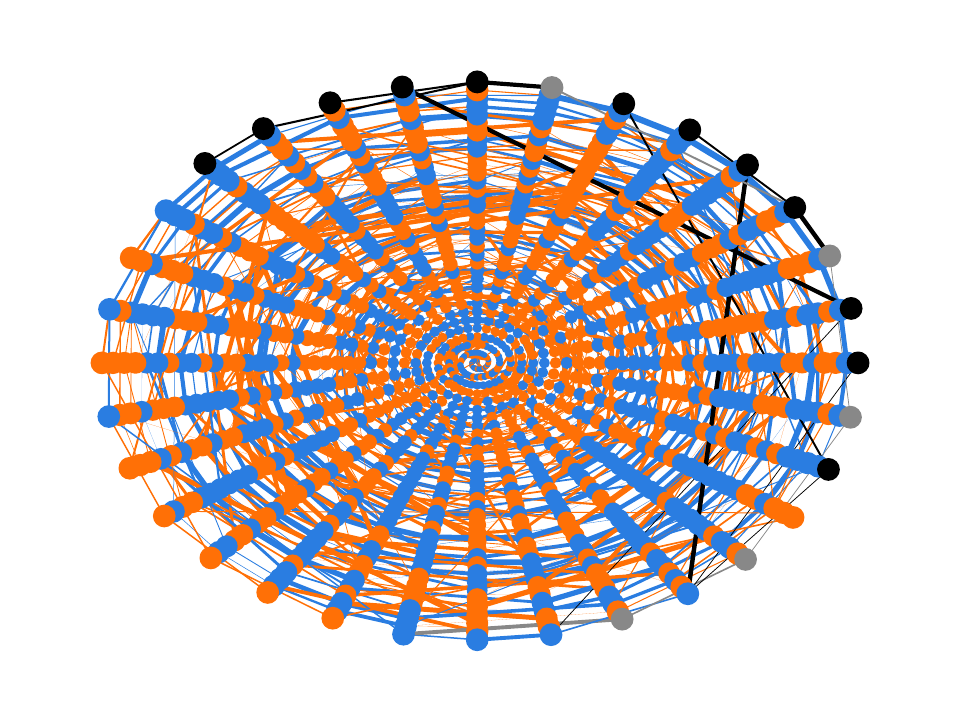} % {0780.png}
\caption{\label{typical-blockchain2} Operation of an example blockchain using four generally-available Advantage and Advantage2 QPUs with $\mathcal{N}_{\rm zeros}=64$ and 100 miners with data collected across more than 2 weeks, with 1041 total block broadcasts and using basic +/-1 Chainwork. The color scheme matches that of Figure \ref{typical-blockchain-14nodes}. Node placement and line-thickness conventions follow Figure \ref{typical-blockchain}.  It is apparent in the larger number of black and gray (unconfirmed) transactions, and in the increased rate of edges spanning the spiral, that this blockchain is of lower efficiency and larger delay than that presented in Figure \ref{typical-blockchain}. This is consistent with larger $\mathcal{N}_{\rm zeros}$ as shown in Figure~\ref{threshold-P} and \ref{threshold-P2}. Where a solver became unavailable mining and verification was redistributed on available solvers; the only significant disruption of four-solver availability was a four-hour outage on April 15th owing to a severed internet cable at the Information Sciences Institute in California (Advantage\_system6.4). 
Although the cross validation rate fluctuated through such events, this had only a small impact on the stability of the blockchain and the measured efficiency is well matched to the bootstrapped prediction of Figure~\ref{threshold-P} at 53 \%. }

\end{figure}\begin{figure}
\includegraphics[width=\linewidth]{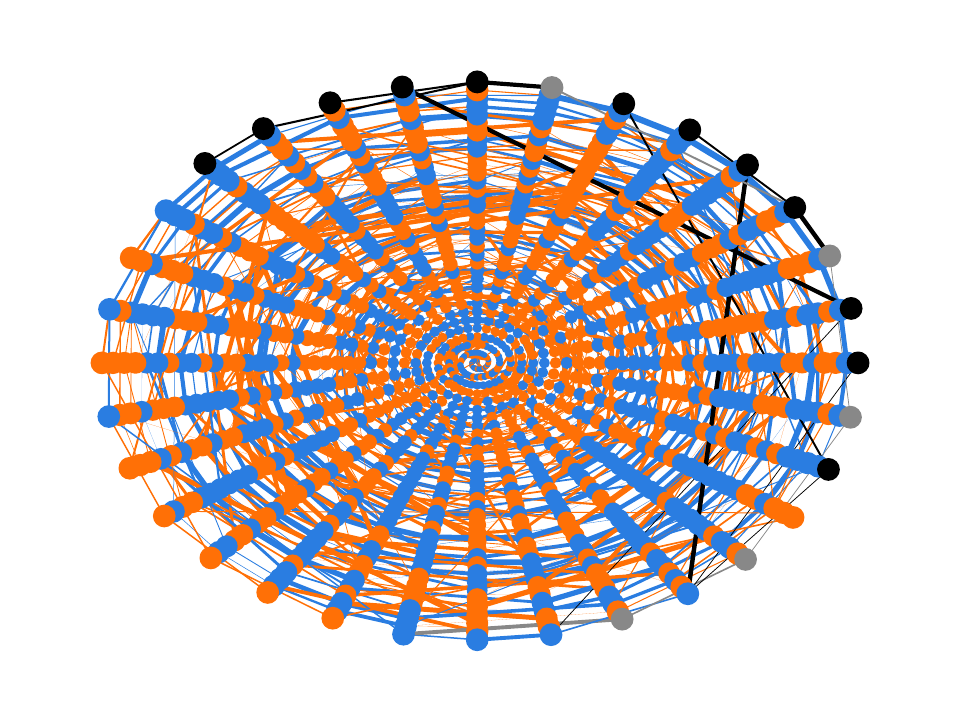}
\caption{\label{typical-blockchain3} Operation of an example blockchain using the Advantage2 QPU and the dimerized biclique spin-glass problems with $\mathcal{N}_{\rm zeros}=64$ and 100 miners, 1101 block broadcasts and using basic +/-1 Chainwork. Like Figure \ref{typical-blockchain2}, the efficiency is estimated at 53\%, with data collected across more than 2 weeks. Despite differences in parameterization of the cubic and biclique demonstrations the median cross-validation rate is comparable, consistent with the matched efficiency. The color scheme, node placement and line thicknesses match the conventions of Figures \ref{typical-blockchain-14nodes} and \ref{typical-blockchain}}
\end{figure}

\begin{figure}
\includegraphics[width=\linewidth]{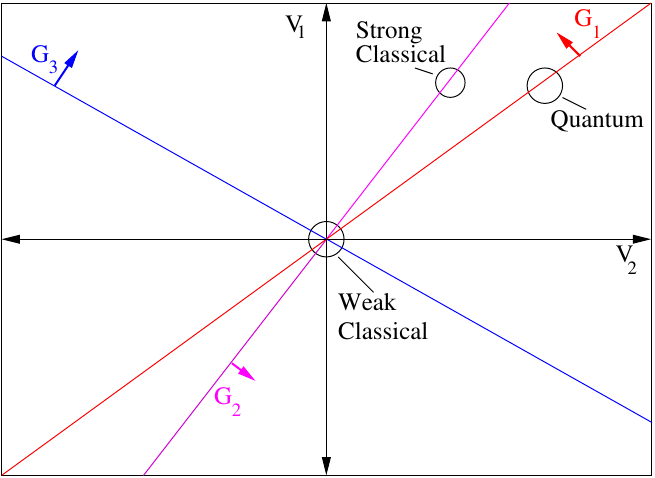}
\caption{\label{hyperplane projection} A high-dimensional statistics vector $V$ is estimated; here only 2 dimensions are shown. Quantum (and classical) estimators are subject to some uncertainty represented as a ball, in high dimensions and with reasonable variance in estimation, all estimators become well resolved in space. By standard locality-sensitive hashing we can digitalize the result. We record what side of each hyperplane the statistics are on. With respect to hyperplanes $G_{2}$ and $G_{3}$ quantum statistics are encoded as +1 and +1, but the outcome with respect to $G_{3}$ is subject to experimental uncertainty. A weak classical estimator, such as any estimator without access to $J$, produces uniform random results. Stronger estimators might produce outcomes correlated in space with the quantum result. Hyperplane distributions can be adjusted by displacement of the origin, or selecting hyperplanes orthogonal to strong classical estimators. $G_{2}$ (a projection orthogonal to the strong classical estimator) allows high confidence in the quantum result, but both the strong and weak classical estimators shown produce outcomes uncorrelated with the quantum result.}
\end{figure}

\begin{figure}
\includegraphics[width=\linewidth]{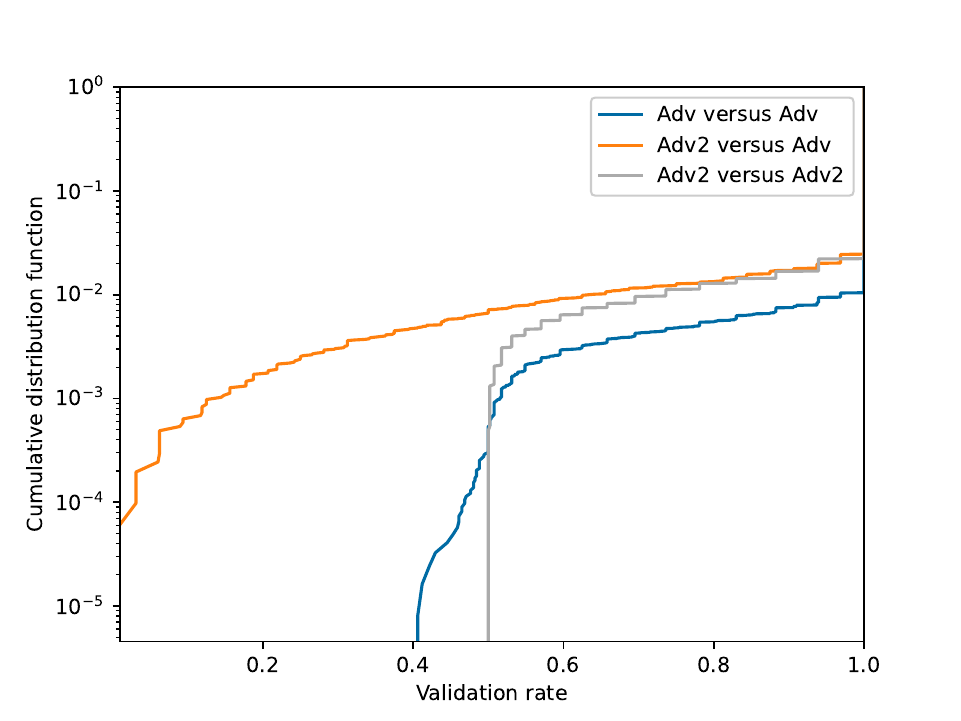}\\
\includegraphics[width=\linewidth]{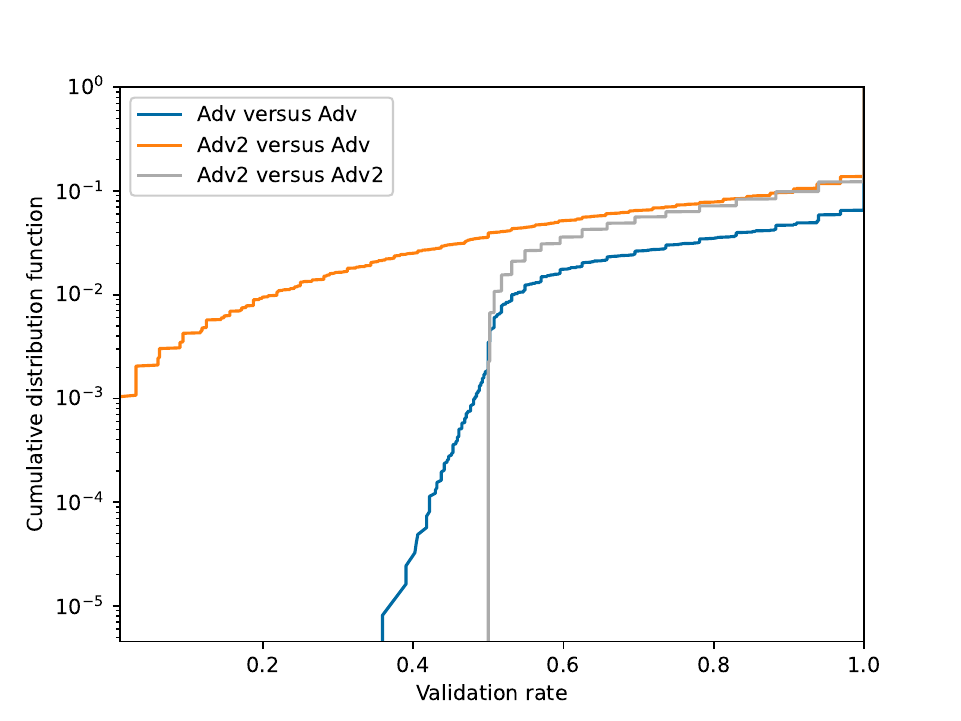}
\caption{\label{Bit-error-rates} Validation rates with respect to experiments on different QPUs.
Validation rates are higher using a common QPU, or a common QPU architecture (Advantage (Adv) or Advantage2 prototype (Adv2)). They are slightly higher on Advantage QPUs due the smaller size of the prototype Advantage2 prototype QPU combined with the bound of 1 second of QPU access time.
On each QPU the probability of an outcome 1 is measured on each QPU from $32$ independent (with randomized embeddings) programmings, for 20 $J$ realizations and total of $20480$ projections. From these $4 \times 20 \times 32$ programmings we infer bit error rates for different QPU combinations. Three Advantage QPUs are used, the relatively small variability between these QPUs is not shown.  (Top) Hyperplanes are chosen with independent and normally distributed components. (Bottom) With the same hyperplanes projected into the space orthogonal to $J$ (see Section \ref{HyperplaneDistributionMitigationsSection}).}
\end{figure}

\begin{figure}
\includegraphics[width=\linewidth]{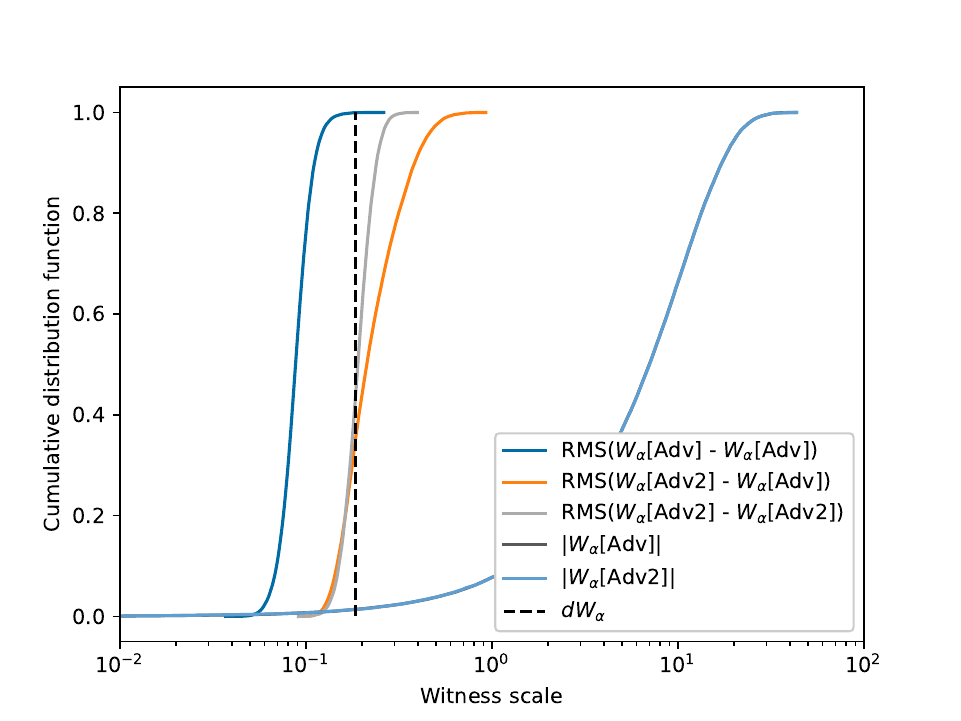}\\
\includegraphics[width=\linewidth]{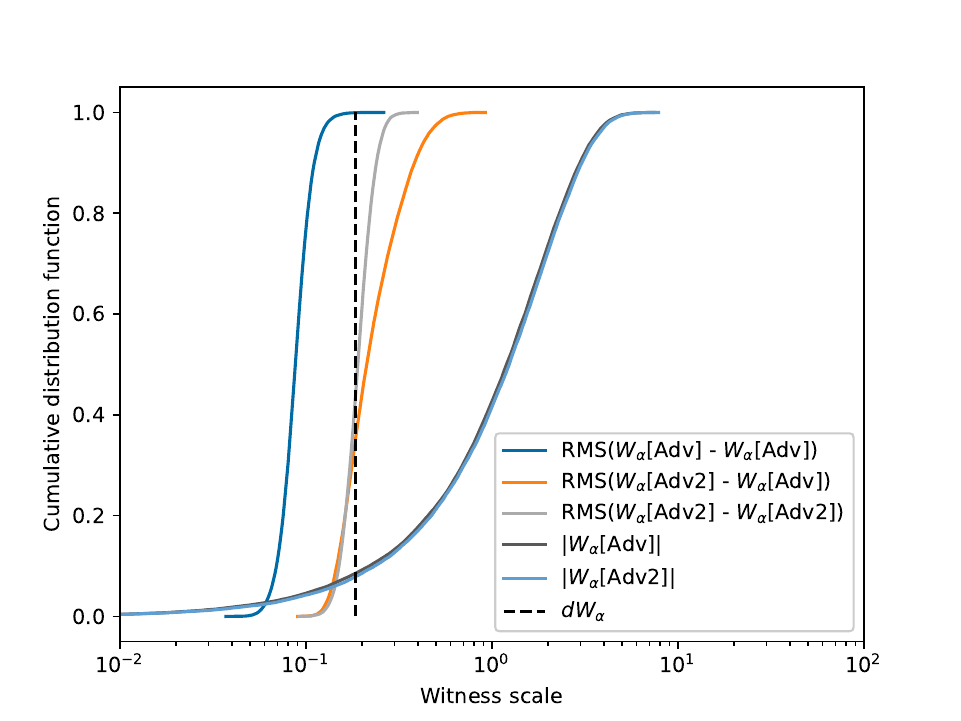}
\caption{\label{dWalpha} Witnesses $W_\alpha$ measured on Advantage (Adv) and Advantage2 (Adv2) QPUs are similar, and much larger than the root mean square (RMS) difference between $W_\alpha$ in different quantum-hash calculations (programmings with independent embeddings on a fixed system, or between systems, as labelled). $\delta W_\alpha$ is shown as a root mean square value across all programmings (variation of embedding and QPUs). (Top) Each hyperplane has independent and normally distributed components. (Bottom) With the same hyperplanes projected into the space orthogonal to $J$. This enhances spoof-resistance, but reduces the signal to noise ratio ($W_\alpha$ are made smaller relative to $d W_\alpha$). (Top) Hyperplane are chosen with independent and normally distributed components. (Bottom) With the same hyperplanes projected into the space orthogonal to $J$ (see Section \ref{HyperplaneDistributionMitigationsSection}).} 
\end{figure}

\begin{figure}
\includegraphics[width=\linewidth]{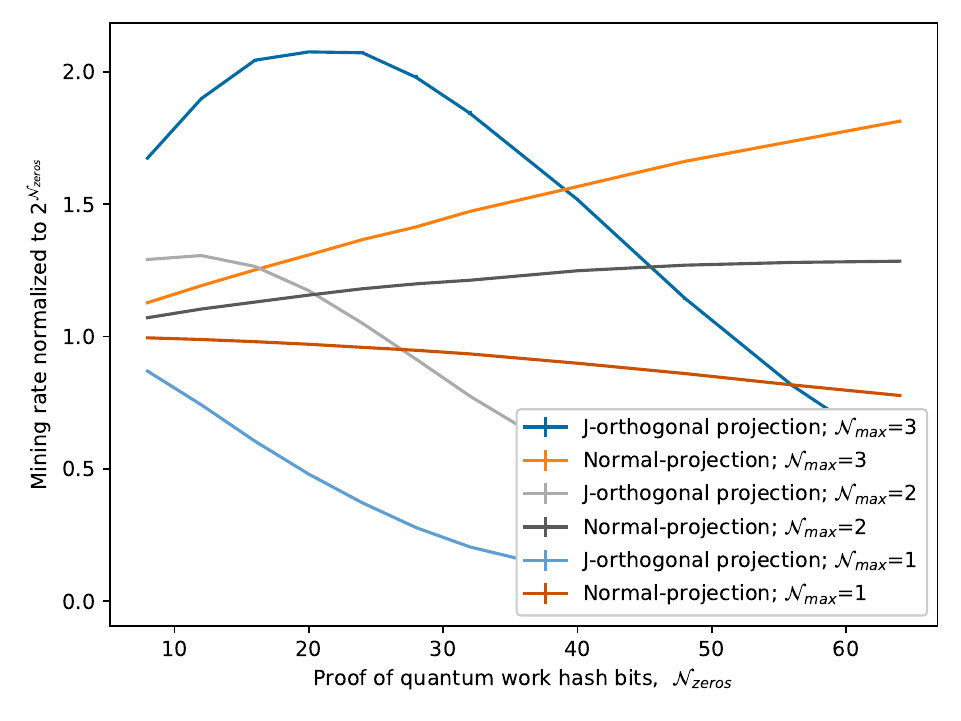}
\caption{\label{MiningRate} Expected hashes required per broadcast with confidence-based Chainwork measured by Monte Carlo sampling. For basic +/- Chainwork the expected number of hashes required per broadcast is precisely $2^{\mathcal{N}_{\rm zeros}}$; confidence-based Chainwork is close to this, with corrections shown. As $\mathcal{N}_{\rm max}$ increases, blocks are easier to mine. As $\mathcal{N}_{\rm max}$ increases, it is easier to find work-satisfying hashes. Using J-orthogonal hyperplanes, more bits are uncertain and it becomes harder to find a confidence-satisfying bit sequence (mining rate goes down).}
\end{figure}

\begin{figure}
\includegraphics[width=\linewidth]{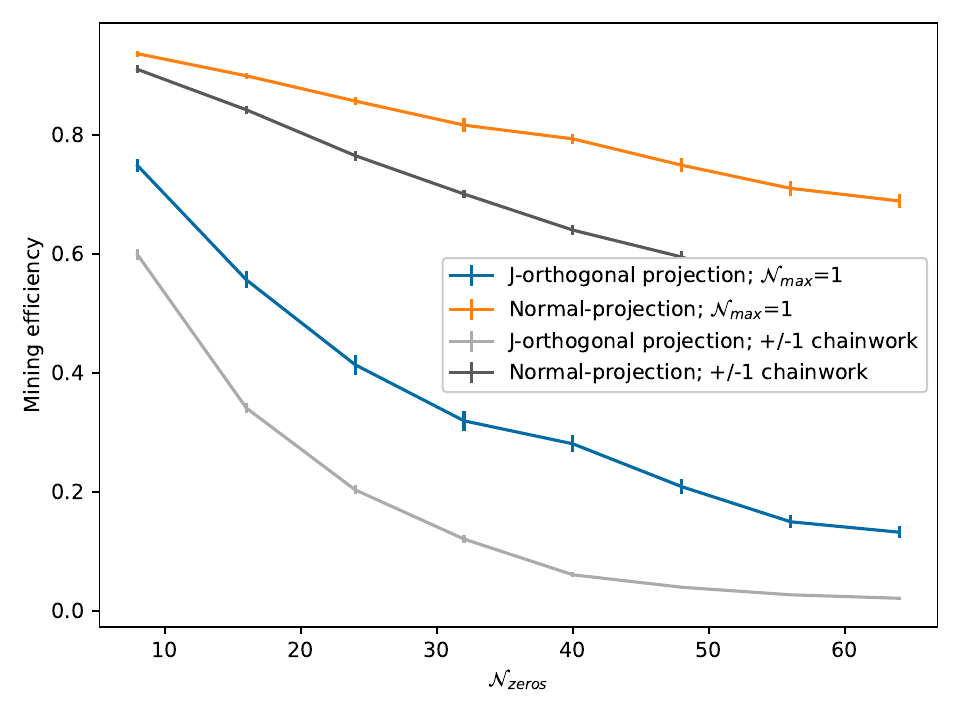}\\
\includegraphics[width=\linewidth]{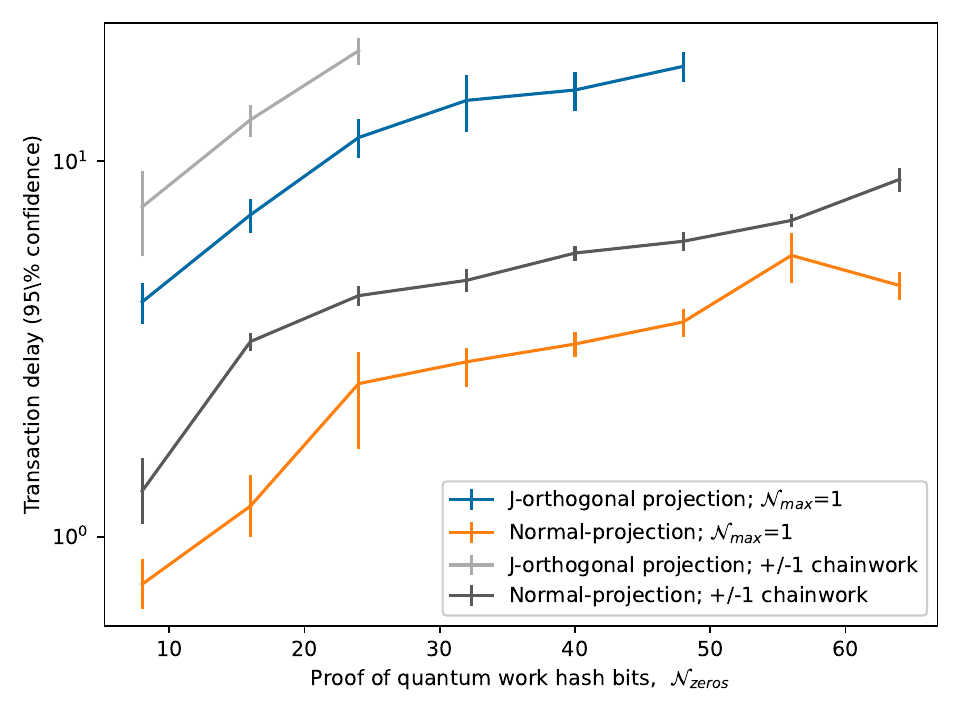}
\caption{\label{threshold-P2} Statistics from 16 chains, each of length 512-1024 using the QPU-witness bootstrapping methodology (see Appendices \ref{StatisticsSection} and \ref{ResamplingWitnessesSection}). Hyperplanes are either i.i.d normally distributed or are restricted to be orthogonal to $J$ (strengthening hashes against weak classical estimators). Resistance to spoofing increases uncertainty in validation, reducing efficiency. The Chainwork uses either the basic +/-1 definition or the confidence-weighted definition with $\mathcal{N}_{\rm max}=1$ or $2$: confidence-weighted strongest chains are more efficient. Statistics are established via the bootstrapping method, with statistic in agreement with the full experiments shown in Figures \ref{typical-blockchain}, \ref{fig:confidenceworkPlusJorthog} and \ref{typical-blockchain2}. (Top) Mining efficiency measures the fraction of block broadcasts entering the strongest chain in the limit of many miners. (Bottom) Transactions (blocks) are uncertain until agreed upon by the majority of miners. Transaction delay with 95\% confidence measures the depth required for a consolidation of 95\% of mining power in the limit of many miners. Stakeholders might consider blocks buried this deep within their strongest chain as immutable with 95\% confidence; higher confidence implies higher delays. Data is truncated where mining efficiency drops below 0.2, as mining depth becomes strongly biased as a function of the finite blockchain size.}
\end{figure}

The dimerized cubic lattice is equivalent to a cubic lattice minor-embedded in the QPU with chain strength 1, with a regular pattern of connectivity between chains. All couplers are $\pm 1$ uniformly and independently distributed. This matches the cubic-nodimer ensemble of \cite{supremacy}. A dimerized biclique is equivalent to an embedded biclique ($K_{n,n}$) model. Couplers are $\pm 1$ on chains, and $\pm \frac{1}{\sqrt{n}}$ between chains in a regular pattern. For the dimerized biclique, the regular pattern of inter-chain couplers is specific to the Advantage2 QPU, so we do not perform mining or cross-QPU validation on Advantage QPUs. The dimerized bicliques ensemble matches that of \cite{supremacy}.

For 3D spin-glasses quantum-hash computation (for mining or validation stages) on each of the four generally-available QPUs, in each case with parallel embeddings (because we consider relatively small problems of 128 qubits, we can program in parallel many copies). We fix the annealing time on the Advantage2 device to be \SI{5}{ns}, and fix annealing times for the other three Advantage devices close to \SI{10}{ns}, so as to approximately maximize the cross-platform validation rate (a median with respect to a test set of spin-glass disorder realizations and hyperplanes).
Figure \ref{collapse} shows rescaling factors necessary for a collapse of the ensemble-average energy. These ratios approximately minimize the correlation error, and maximize the validation rate.
The difference in annealing times reflects a difference in energy scales between devices. The unitless unitary evolution then becomes well matched, at least throughout the critical region~\cite{supremacy}. Energy scales are shown with respect to the Hamiltonian \eqref{HS} in Fig.~\ref{schedules}. We then fix the number of reads to fully utilize the maximum QPU access time (\SI{1}{second}) per programming, and otherwise use default parameters. For the biclique problem we have 72 qubits and fix the Advantage2 annealing time to \SI{15}{ns}. Demonstrations plots relate to the cubic lattice ensemble, unless stated otherwise.

Per quantum hash computation (mining or validation) we choose uniformly at random among the available QPUs and then embed the problem subject to an automorphism and spin-reversal transform on each available embedding. The effect of these transformations is to randomize the mapped position and signs on programmed fields and couplers, in effect emulating the enhanced noise that might be anticipated from (more) distributed computation.

Returned sample sets are processed to nearest-neighbor correlations $C$ (1 per programmed coupler), that are then projected with independent and identically distributed hyperplanes to determine witnesses. Hyperplanes have independent and normally distributed components. In the case of Fig.~\ref{threshold-P2} we also consider a variation in which hyperplanes are projected into the space  orthogonal to $J$, as described in \ref{HyperplaneDistributionMitigationsSection}. $C$ is correlated with $-J$, and we remove the linear part of this correlation with such a projection, making the quantum statistics harder to spoof classically. A visualization of the projection process is shown in Fig.~\ref{hyperplane projection}, distributions for witnesses are shown in Fig.~\ref{dWalpha}. Uncertainty in witness values is typically much smaller than the witness value. Witness values vary more between Advantage and Advantage2 QPUs, than for same-generation QPU validations.

Witnesses are thresholded to determine hashes. This digitalization of the correlations is the standard locality-sensitive hashing method, which guarantees preservation of local distances. With enough projections a beyond-classical (small) error in correlations translates to a small Hamming distance in the resulting bit sequence. The probability that bits are validated (reproduced) under a repeated experiment are shown in Fig.~\ref{Bit-error-rates}. More than 99\% of bits are typically validated, but the remaining bits can have high uncertainty, particularly with respect to cross-processor values.

We performed simulations of many chains with 100 miners modelling a mining pool and transactions, or taking the limit of many miners (where no miner mines more than one block). Examples are shown in Figs.~\ref{typical-blockchain}, \ref{typical-blockchain2}, \ref{typical-blockchain3} and \ref{fig:confidenceworkPlusJorthog}. However, only the distribution of witness statistics impacts the structure and stability of the chain, assuming no simultaneous broadcasts, via mining events and stakeholder validation. Understanding this we were able to accelerate mining (Appendix \ref{MiningRateSection}), and simulate quantum hashes by bootstrapping (Section \ref{ResamplingWitnessesSection}). We first describe our main statistical measures of blockchain stability in Section \ref{StatisticsSection}. 

\subsection{Blockchain Efficiency and Transaction Delay}
\label{StatisticsSection}
For a blockchain to be successful, (1) efficiency must be high so mining work is reliably rewarded and (2) delays in transaction confirmation must be small to enable use as a real-time ledger. These quantities are strongly correlated but subtly different.

Any stakeholder identifies a node with maximum Chainwork that defines the strongest chain; this can be considered a random variable and defines a unique strongest path from the genesis node. 

We define blockchain efficiency as the expected number of blocks in the strongest chain (path) divided by the total number of blocks broadcast, in the limit of many blocks. For viable parameterizations this concentrates at a finite value as the number of block broadcasts becomes large. A per-blockchain realization value can be estimated by averaging values obtained between broadcast (mining event) $512$ and $1024$. Figs.~\ref{threshold-P} and \ref{threshold-P2} show the mean and standard deviation with respect to $16$ independent blockchains.

We can establish two such independent strongest paths, using two stakeholders who have not mined any blocks, and define their pair delay, $D_{pair}$, as the number of blocks in the first path absent from the second path (i.e. the depth at which there is disagreement in the strongest chain). If two users agree on the strongest chain this delay is $0$, and is otherwise positive. Stakeholders can be miners: the pair delay describes the distribution of mining power relative to a typical user. If pair-delays are with high probability less than $\mathcal{D}$, then a stakeholder can have corresponding confidence that blocks of depth at least $\mathcal{D}$ on the strongest chain are immutable, since mining power is concentrated on branches compatible with the block validity. We call $\mathcal{D}$ the transaction delay.

For figure \ref{threshold-P2} we measure 1 pair-delay per chain following each mining event on chains of length from 512 up to 1024. We determine the empirical distribution of $D_{pair}$, discarding statistics from the first half of each chain. We threshold the distribution to determine a transaction delay with 95\% confidence, presenting the mean and variance with respect to 16 blockchains as indicated in captions. Although pair-delays measured this way are correlated, for each data point we have at least thousands of effectively independent samples, so that a 95\% quantile is established with reasonable confidence~\cite{RevisitingGelmanRubin}. 

\subsection{Mining Acceleration for Blockchain Analysis}
\label{MiningRateSection}
Each mining event requires $O(2^{\mathcal{N}_{\rm zeros}})$ nonce attempts (quantum computations). For purposes of blockchain structural analysis we show methods whereby only $O(1)$ nonces are required per mining event. This allows us to present,  with practical compute resources, results on stability for large $\mathcal{N}_{\rm zeros}$, limited primarily by the number of validating computations rather than the exponentially growing mining timescale.

A valid mining event occurs when the QPU produces witnesses that (a) after thresholding contain $\mathcal{N}_{\rm zeros}$ zeros (with the +/-1 Chainwork definition) or (b) satisfy the confidence-based Chainwork criteria. 

We first make the approximation of perfect network synchronization, discounting the rare event of two miners simultaneously completing a proof-of-work. This approximation is reasonable if the mining rate is small enough.

We employ a hashing method where the overall sign on any hyperplane is uniformly distributed in $\{-1,1\}$. For this reason the probability a nonce describes a broadcastable (valid) block in case (a) is $1/2^{-\mathcal{N}_{\rm zeros}}$; mining is successful if and only if the hyperplane signs $H_\alpha = \mathrm{sign}(W_\alpha)$. We can therefore emulate properly distributed statistics of a fair mining event by first calculating $W_\alpha$ for a uniformly sampled random nonce, and then sampling the hyperplane signs conditioned upon the $\mathcal{N}_{zeros}$ requirement.

For a confidence-based Chainwork, the probability of successful mining is a function of more than the signs on the hyperplanes. To accelerate mining events in this case, we compute a large number of quantum hashes to determine many copies of $W_\alpha$ for $\alpha = 1$ to $\mathcal{N}_{\rm zeros}$. In each case, we exhaustively enumerate all hyperplane signs consistent with a mining event. We thereby enumerate a subset of all mining events. These mining events are correlated (in some cases differing only by the signs on hyperplanes), but this is judged to be a weak correlation. Most often only one value for the signs is supported, and events that allow many sign values are rare for the parameter ranges presented. The rate of mining can be inferred, and the mining subset can be used in a randomized order to model the mining.

The probability of confidence-threshold satisfying events is a function of $\delta W$,  $\mathcal{N}_{\rm max}$ and $\mathcal{N}_{\rm zeros}$, and remains approximately equal to $2^{-\mathcal{N}_{\rm zeros}}$ subject to a small correction, as shown in Fig.~\ref{MiningRate}. 

\subsection{Bootstrapping Witness Methodology}
\label{ResamplingWitnessesSection}

For the case that Chainwork is defined as +/-1 for matching (unmatching) sequences, knowing the distribution of the $\mathcal{N}_{\rm zeros}$ bits for each device used in mining and validation is sufficient to evaluate the distribution of Chainwork and thereby the evolution of the blockchain. For the case of confidence-based Chainwork, it is sufficient to know the distribution of $\{W_\alpha\}$. 

For a training set of $20$ nonce values (unitary evolution parameterizations) and $1024$ random projections, the distribution of bits (thence validation rates), and distribution of $W_\alpha$ (thence confidence in validation) can be established as shown in Figs.~\ref{Bit-error-rates} and \ref{dWalpha}. We can fit the witness distributions by Gaussians, and thresholded bit outcomes by Bernoulli distributions. For purposes of accelerating analysis we can then resample these distributions uniformly at random to model a larger set of quantum hashes without use of the QPU experiment each time.

We employ this procedure in Figs.~\ref{threshold-P} and \ref{threshold-P2} with results consistent with full blockchain demonstrations involving finite-number of miners, such as those in Fig.~\ref{typical-blockchain2}. We consider 20480 combinations of model (spin-glass realization) and hyperplanes, for each of the 4 generally-available QPUs. Using 32 programmings, we determine the distribution (Gaussian or Bernoulli for witnesses and bits respectively). We can then use these parameterized distributions to generate reasonable witness (and thresholded witness) distributions. For each mining attempt, we uniformly randomly sample a combination of model and hyperplanes, and then for any mining or validation event we choose uniformly at random a QPU and sample from the corresponding distribution.

We can further save on computation costs by only mining (validating) as necessary for purposes of constructing the blockchain. We need only understand a miner's verification pattern if they are responsible for broadcasting the next block. We therefore construct on-the-fly a miner's Chainwork and establish their strongest chain at the time of mining. With this approximation we can work in the informative limit of well-distributed compute power (many miners), where each miner mines at most one block in any blockchain. The number of validation events required to realize a chain with $L$ blocks broadcast in total is $L(L-1)/2$.

Qualitative justification for this approach relies on the observation that the spin-glass ensemble of unitary evolutions in our demonstration is {\it self-averaging}. The distributional properties of such witnesses fluctuate little between typical spin-glass disorder realizations. Variation in witnesses values between QPUs (and embeddings) appears well described by Gaussian distributions, and since hyperplanes are nearly orthogonal the correlations in their fluctuations are weak. We plot in Figs.~\ref{threshold-P} both results obtained by bootstrapping and those obtained in demonstrations (either with fixed number of miners, or many miners) with good agreement, validating the approximation. Full simulations (every verification and mining event uses a QPU) are used for all other figures.

 \end{document}